\documentclass[a4paper,fleqn]{cas-sc}
\ExplSyntaxOn
\cs_gset:Npn \__first_footerline:
  { \group_begin: \small \sffamily \__short_authors: \group_end: }
\ExplSyntaxOff
\usepackage[authoryear]{natbib}
\def\tsc#1{\csdef{#1}{\textsc{\lowercase{#1}}\xspace}}
\tsc{WGM}
\tsc{QE}
\tsc{EP}
\tsc{PMS}
\tsc{BEC}
\tsc{DE}
\usepackage{subcaption}
\usepackage{caption}
\usepackage{amsmath,amsopn}
\usepackage{physics}
\usepackage{xparse}
\usepackage{multirow}
\usepackage[graphicx]{realboxes}
\usepackage[switch]{lineno}

\begin{document}
%\linenumbers
\let\WriteBookmarks\relax
\def\floatpagepagefraction{1}
\def\textpagefraction{.001}
\shorttitle{}
\shortauthors{M. Rufino, A.L. Lixandrão-Filho, S. Guedes}

\title [mode = title]{A reappraisal of the principle of equivalent time based on physicochemical methods}

\author[1]{M. Rufino}[type=editor,
                        auid=000,bioid=1,
                        prefix=,
                        role=,
                        orcid=0000-0003-4871-5120]
\ead{rufino@ifi.unicamp.br}

\credit{Conceptualization of this study, Methodology, Software}

\address[1]{Departamento de Raios Cósmicos e Cronologia, Grupo de Cronologia, Instituto de Física ``Gleb Wataghin", Universidade Estadual de Campinas, R. Sérgio Buarque de Holanda, 777 - Cidade Universitária, Campinas - SP, 13083-859, Brazil}

\author[1]{A.L. Lixandrão-Filho}[type=editor,
                        auid=000,bioid=1,
                        prefix=,
                        role=,
                        orcid=0000-0002-8343-8942]

\ead{allfilho@ifi.unicamp.br}

\author[1]{S. Guedes}[type=editor,
                        auid=000,bioid=1,
                        prefix=,
                        role=,
                        orcid=0000-0002-7753-8584]

\cormark[1]
\ead{sguedes@ifi.unicamp.br}
\cortext[cor1]{Principal corresponding author}

\begin{abstract}
The main feature of the Fission-Track Thermochronology is its ability to infer the thermal histories of mineral samples in regions of interest for geological studies. The ingredients that make the thermal history inference possible are the annealing models, which capture the annealing kinetics of fission tracks for isothermal heating experiments, and the Principle of Equivalent Time (PET), which allows the application of the annealing models to variable temperatures. It turns out that the PET only applies to specific types of annealing models describing single activation energy annealing mechanisms (parallel models). However, the PET has been extensively applied to models related to multiple activation energy mechanisms (fanning models). This procedure is an approximation that has been overlooked due to the lack of a suitable alternative. To deal with this difficult, a formalism, based on physicochemical techniques, that allows to quantify the effects of annealing on the fission tracks for variable temperatures, is developed. It is independent of the annealing mechanism and, therefore, is applicable to any annealing model. In the cases in which the PET is valid, parallel models, the proposed method and the PET predict the same degrees of annealing. However, deviations appear when the methods are applied to the fanning models, with the PET underestimating annealing effects. The consequences for the inference of thermal histories are discussed.
\end{abstract}

%\item A framework, based on physicochemical techniques, capable of calculating the reduced length of fission tracks in a variable thermal history, was developed to be compared with the principle of equivalent times.
%\item Predictions of track length shortening, after a linear cooling, are the same for both methods when the single activation energy (parallel) annealing models are considered.
%\item Predictions of track length shortening, after a linear cooling, deviate for both methods when the multiple activation energy (fanning) annealing models are considered.
%\item The Principle of Equivalent time applies to parallel models but is an approximation to fanning annealing models, which implies in additional uncertainty in thermal history inference.


%\ExplSyntaxOn \cs_gset:Npn \__first_footerline: { \group_begin: \small \sffamily \__short_authors: \group_end: } \ExplSyntaxOff tirar o preprint Elsevier
\maketitle
\section{Introduction}
\label{Intro}
The Principle of Equivalent Time (PET) is one of the basic ingredients for the inference of thermal histories in Fission-Track Thermochronology (FTT). The PET states that the rate of track shortening due to temperature is independent of its previous thermal history. Thus, any thermal history may be replaced with a constant temperature heating for an equivalent time resulting in the current fission-track length. Since the models that constrain the annealing kinetics are only applicable to constant temperature heating events, it is the PET that allows for the inference of variable thermal histories from the fission-track age and from the distribution of fission-track lengths measured in the sample.

The PET was first proposed by \citep{goswami1984quantitative}. Later on, \citet{duddy1988thermal} established a practical method of finding thermal histories, applying the PET, that has been mostly unchanged since then. They also demonstrated that the PET is only valid for the single activation energy Arrhenius annealing equation. Such equation is represented as parallel straight isoretention (same fission-track length) curves on the pseudo-Arrhenius space (logarithm of time as a function of inverse temperature, Fig.~\ref{fig:pseudoArrheniusLinear}) and is called Parallel Arrhenius equation. However, they applied the PET to the Fanning Arrhenius (FA) model (Laslett et al., 1987), in which the isoretention curves diverge from a single point with different slopes, implying different activation energies. They recognized that this procedure is an approximation, since the PET only applies to parallel models, but argued that their fanning model deviated only slightly from a parallel one.

Annealing models continued to evolve. \citet{Carlson1990model} presented a modified version of the parallel model (Fig.~\ref{fig:pseudoArrheniusLinear}). \citet{crowley1991Arrhenius} proposed versions of the Parallel and Fanning equations that are curved in the pseudo-Arrhenius space (Fig.~\ref{fig:pseudoArrheniusCurvilinear}), implying activation energies that vary with the temperature of annealing. The Fanning Curvilinear (FC) model has been shown to produce better geological extrapolations than the other models for the apatite \citep{ketcham2007improved} and zircon \citep{guedes2013improved} fission-track systems. It is currently the model of choice in most geological studies using the FTT \citep{Ketcham2019} in thermal history codes relying on the PET to apply the annealing equations for the inference of thermal histories \citep{ketcham2005software, gallager2012software}. The joint application of the PET and FC equation is an approximation that has been overlooked due to the lack of an alternative to deal with the variable temperature thermal histories.

Recently, \citet{rufino2022arrhenius} applied a physicochemical technique to the Fission-Tack Arrhenius equations and were able to formulate the annealing kinetics in terms of the reaction rate constant, which is the fundamental quantity related to the activation energy \citep{Arrhenius1889, cohen2007iupac}. The reaction for the annealing process is the recombination of displaced atoms and vacant sites that form the track. Once the rate constant is determined for the annealing equations, they can be represented in the Arrhenius space (logarithm of the rate constant as a function of the inverse temperature) and their trends can be used to retrieve the general mechanisms underlying the Arrhenius models. The rate constant encodes the most fundamental features of annealing. Once it is determined, the shortening of the fission tracks may in principle be quantified not only for constant temperatures but also for varying ones, being an alternative to the PET, without the single activation energy restriction.

The Arrhenius annealing equations can be derived from the rate constant for the case of constant temperature annealing \citep{rufino2022arrhenius}. The next step is to apply the physicochemical approach to the variable temperature annealing of the fission tracks and compare the results to the ones obtained with the PET. The length shortening of fission tracks is calculated for cooling temperature-time (T-t) paths with different slopes using the parallel, fanning and Carlson models as they are the representations of different activation energy mechanisms. The temperature indexes Closure and Total Annealing temperatures, calculated using the PET and the rate constant techniques, are presented and compared to illustrate the differences between both approaches for the geological extrapolation.

\section{Method}
\label{Method}
\subsection{The physicochemical perspective of fission track annealing}
The kinetics of chemical reactions can be described by the Arrhenius equation \citep{Arrhenius1889}, which relates the temperature derivative of the reaction rate $k$, the universal gas constant $R$ and a constant $q$, related to a change in the standard internal energy \cite[p.494]{Laidler1984Arrhenius}:

\begin{equation}\label{eq:difArr}
	\dv{\ln k(T)}{T} = \frac{q}{RT^2}.
\end{equation}

Eq. (\ref{eq:difArr}) can be solved for the reaction rate as a function of temperature $k(T)$ using a pre-exponential factor $A$ and the Arrhenius activation energy $E_a$:

\begin{equation}\label{eq:arrk}
	k(T) = A\exp\left(-E_a/RT\right).
\end{equation}

Chemical processes that obey Eq.~\eqref{eq:arrk} result in straight lines with slope $-E_a/R$ in Arrhenius plots ($\ln k \times 1/T$). Among the fission-track annealing models, the Parallel Arrhenius is the only one that actually fits this formulation of a single constant activation energy. Deviations from Eq.~\eqref{eq:arrk} are quite common. To enable a more complete study of chemical reactions, the International Union of Pure and Applied Chemistry (IUPAC) has defined the Arrhenius activation energy \citep{cohen2007iupac}:

\begin{equation}\label{eq:Ea_definition}
	E_a = - R \dv{\ln(k)}{(1/T)}.
\end{equation}

The Arrhenius activation energy, as defined by Eq.~\eqref{eq:Ea_definition}, is an empirical quantity aimed to be a kinetic parameter that can vary with the temperature of the reaction medium. Its determination depends on the previous knowledge of the rate constant, the quantity that encodes the reaction kinetics. Thus, for application to the fission-track system, the reaction rate constant associated with the annealing mechanisms must be found, which can be done using the formalism of studies in solid state processes \citep{Vyazovkin2015Book}.

\begin{figure}[!h]
\centering
    \subfloat[]{\label{fig:pseudoArrheniusLinear}
    \includegraphics[width=0.49\linewidth]{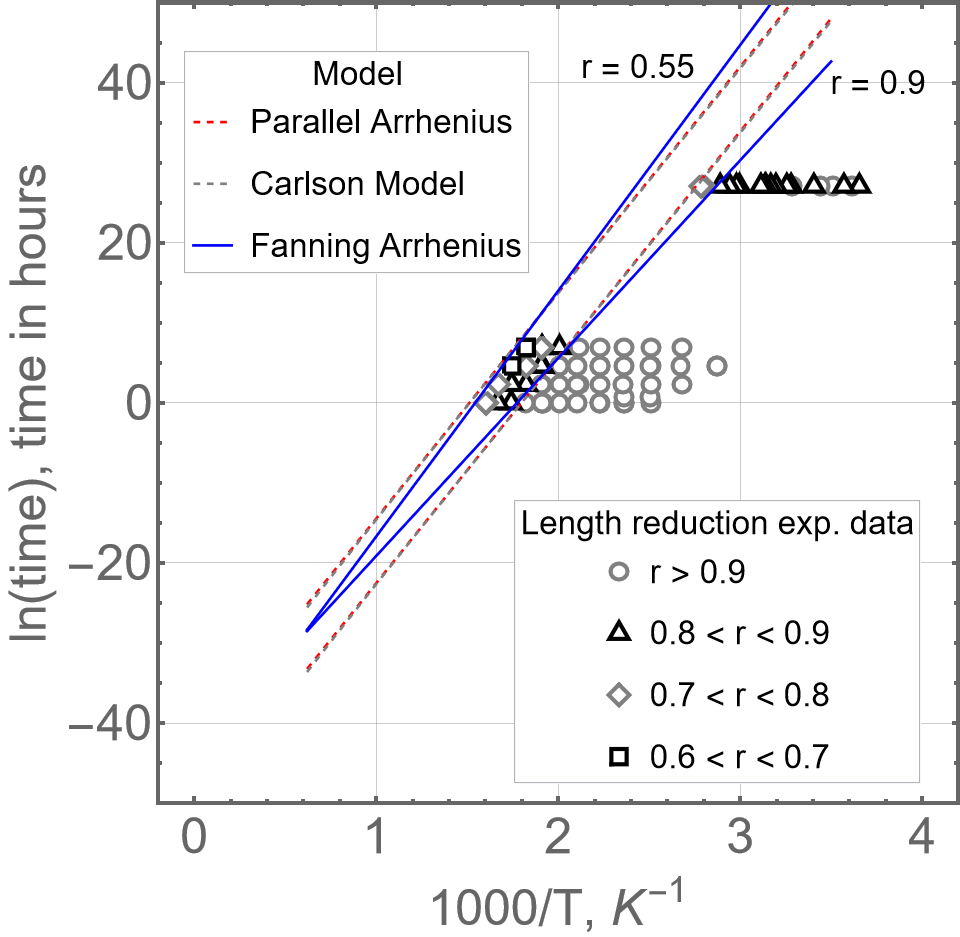}
}
    \subfloat[]{\label{fig:pseudoArrheniusCurvilinear}
    \includegraphics[width=0.49\linewidth]{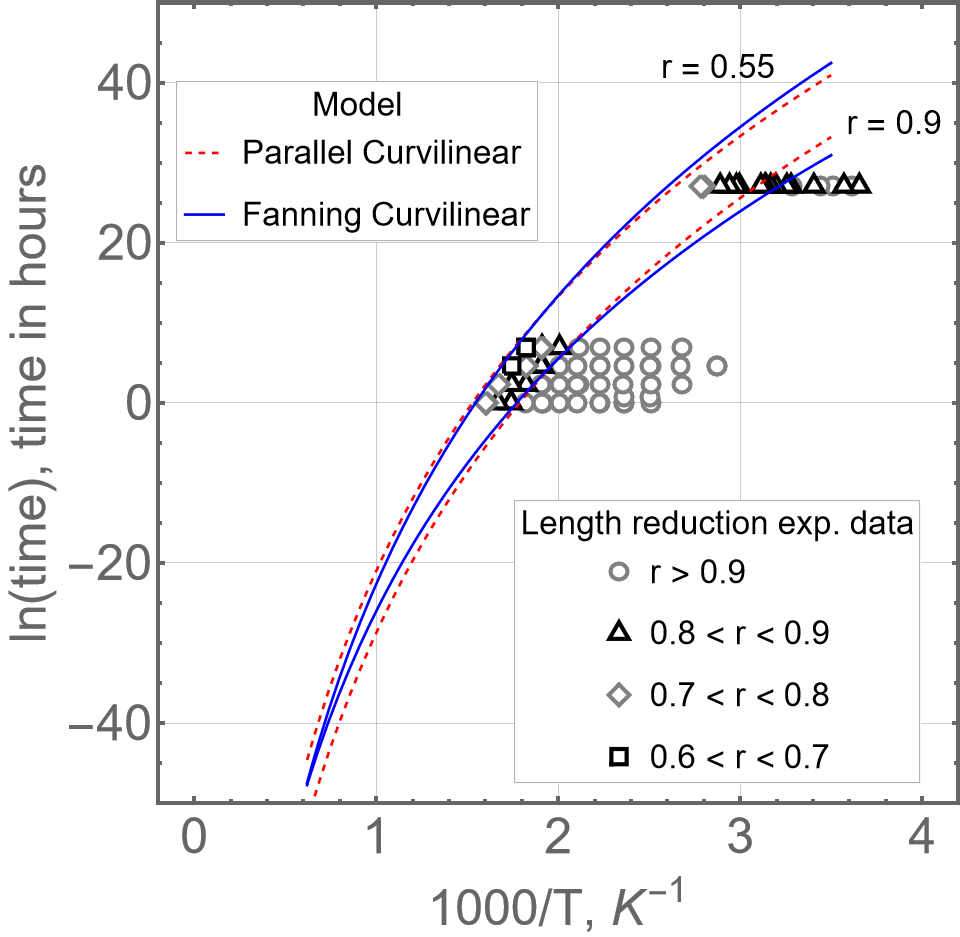}
}
\caption{Representation of the Arrhenius fission-track annealing models in the pseudo-Arrhenius plot. (a) Fanning Arrhenius, Parallel Arrhenius, and Carlson models. (b) Fanning Curvilinear and Parallel Curvilinear models. Laboratory annealing data are c-axis projected reduced fission-track lengths from Durango apatite \citep{Carlson1999data}. Data from the geological benchmark KTB \citep{Wauschkuhn2015KTB} are included only for reference. The models are represented as isoretention curves. Points on these curves are the temperature and time of constant temperature heating resulting in the same reduced length. 
}
\label{fig:pseudoArrhenius}
\end{figure}

The annealing kinetics of fission tracks is described by empirical \citep{laslett1987thermal, crowley1991Arrhenius, LaslettGabraith1996, Rana2021} or semi-empirical \citep{Carlson1990model, guedes2006kinetic, guedes2013improved} equations relating the reduced track length, $r = L/L_0$ (where $L$ is the length of the fission track after heating and $L_0$ is the unannealed fission-track length), with the duration, $t$, of the constant temperature ($T$) heating. The general form of the annealing equations is:

\begin{equation}\label{eq:general annealing equation form}
    g(r) = f(t,T),
\end{equation}

\noindent in which $g(r)$ is a transformation of $r$ and $f(t,T)$ defines the geometrical characteristics of the isoretention curves in the pseudo-Arrhenius space ($\ln t \, \times \, 1/T$, Fig.~\ref{fig:pseudoArrhenius}). The Parallel Arrhenius (PA, Eq.~\eqref{eq:PA}) and Fanning Arrhenius (FA, Eq.~\eqref{eq:FA}) equations \citep{laslett1987thermal}, the Parallel Curvilinear (PC, Eq.~\eqref{eq:PC}) and Fanning Curvilinear (FC, Eq.~\eqref{eq:FC}) models \citep{crowley1991Arrhenius} as well as the Carlson Model (CM, Eq.~\eqref{eq:CM}) that mixes the Parallel Arrhenius and Parallel Curvilinear models in the same equation \citep{Carlson1990model}, are used in this analysis. The transformation function $g(r)=\ln(1-r)$ was chosen because it carries no fitting parameters and was shown to produce good fits to annealing data \citep{guedes2022generalization}. In addition, it arises naturally from the physicochemical formulation of the fission track annealing \citep{rufino2022arrhenius}, as will be shown below. More comprehensive descriptions of the annealing models can be found elsewhere \citep{Carlson1990model, Ketcham2019, guedes2022generalization}. 

The annealing data set on the c-axis projected fission tracks for Durango apatite \citep{Carlson1999data} was used for model fitting. Durango apatite annealing data was chosen because Durango is a well-known standard sample often used in methodological studies \citep{Green1986, Carlson1990model, Ketcham1999, ketcham2007improved, Rana2021, guedes2022generalization, rufino2022arrhenius}. The fitting parameters for PA, PC, FA and FC models are the same presented in \citet{rufino2022arrhenius}. They were numerically determined using the function \texttt{nlsLM} of the package \texttt{minpack.lm} \citep{Elzhov2016} written in \texttt{R} language, which applies the Levenberg–
Marquardt algorithm to minimize the residual sum of square (RSS), using the squared inverse of $r$ uncertainties as weights. With the same method, fitting parameters were also obtained for the CM model. The fitting parameters are presented in the last column of Table \ref{tab:synthesis}.

\begin{table*}[]

\caption{Effective reaction rate constant, $r$ reduction and equivalent time equations associated with the fission-track annealing models}
\resizebox{\textwidth}{!}{
\setlength\abovedisplayskip{2pt}
\setlength\belowdisplayskip{2pt}
\begin{tabular}{lcl}
\hline
FT models &  Equations    & Parameters (standard error)   \\ \hline
\begin{tabular}[c]{@{}c@{}} Parallel Arrhenius \\ (PA) \end{tabular}        & \begin{tabular}[l]{@{}p{9.5cm}@{}}
\begin{equation}   
f_{PA}(t,T)= c_0 + c_1 \ln(t) + \frac{c_2}{R T}  
\tag{PA1} \label{eq:PA}  
\end{equation}
\begin{equation}
        k_{ef}(T)_{PA}=c_1 e^{c_0/c_1}\exp(\frac{c_2/c_1}{R T})
        \tag{PA2}
        \label{eq:kPA}
 \end{equation}
\begin{equation}
         E_a^{PA}=-\frac{c_2}{c_1}
         \tag{PA3}
         \label{eq:EaPA}
\end{equation}
\end{tabular}
    & \begin{tabular}[c]{@{}l@{}} $c_0$ = 5.631 (0.220) \\ $c_1$ = 0.1865 (0.0066) \\ $c_2$ = -10.46 (0.31) kcal/mol\\ $\chi_\nu^2$ = 2.65 \end{tabular} \\ \hline
\begin{tabular}[c]{@{}c@{}} Parallel curvilinear \\ (PC) \end{tabular}       &  \begin{tabular}[l]{@{}p{9.5cm}@{}}
\begin{equation}
        f_{PC}(t,T) = c_0 + c_1 \ln(t) +c_2\ln\left(\frac{1}{R T} \right)
    \tag{PC1}
    \label{eq:PC}
\end{equation}
\begin{equation}
        k_{ef}(T)_{PC}=c_1 e^{c_0/c_1} \left(R T\right)^{-c_2/c_1}
        \tag{PC2}
        \label{eq:kPC}
\end{equation}
\begin{equation}
         E_a^{PC}(T)=-\frac{c_2}{c_1}RT
         \tag{PC3}
         \label{eq:EaPC}
\end{equation}
\end{tabular}
& \begin{tabular}[c]{@{}l@{}} $c_0$ = -4.910 (0.096) \\ $c_1$ = 0.1944 (0.0060) \\ $c_2$ = -9.610 (0.244) \\ $\chi_\nu^2$ = 2.12 \end{tabular}  \\ \hline
\begin{tabular}[c]{@{}c@{}} Carlson Model \\ (CM) \end{tabular}       &  \begin{tabular}[l]{@{}p{9.5cm}@{}}
\begin{equation}
        f_{CM}(t,T) =c_0+c_1\ln(t)+c_1\ln(RT)+\frac{c_2}{RT}
        \tag{CM1}
        \label{eq:CM}
\end{equation}
\begin{equation}
        k_{ef}(T)_{CM}=c_1e^{c_0/c_1}\exp(\frac{c_2/c_1}{RT})RT
        \tag{CM2}
        \label{eq:kCM}
\end{equation}
\begin{equation}
        E_a^{CM}(T)=-\frac{c_2}{c_1}+RT
         \tag{CM3}
         \label{eq:EaCM}
\end{equation}
\end{tabular}
& \begin{tabular}[c]{@{}l@{}} $c_0$ = 5.426 (0.2155) \\ $c_1$ = 0.1867 (0.0066) \\ $c_2$ = -10.25 (0.2994) kcal/mol \\  $\chi_\nu^2$ = 2.63 \end{tabular}  \\ \hline
\begin{tabular}[c]{@{}c@{}} Fanning Arrhenius \\ (FA) \end{tabular}       &  \begin{tabular}[l]{@{}p{9.5cm}@{}}
\begin{equation}
        f_{FA}(t,T) = c_0 + c_1 \frac{\ln(t)-c_2}{\frac{1}{RT}-c_3}
        \tag{FA1}
        \label{eq:FA}
\end{equation}
\begin{equation}
        k_{ef}(t,T)_{FA}=\frac{c_1\exp[(1-n)f_{FA}(t,T)]}{t(1-c_3RT)}
        \tag{FA2}
        \label{eq:kFA}
\end{equation}
\begin{equation}
         E_a^{FA}(t,T)=\frac{(R T)^2 \left[c_1(n-1) (c_2 - \ln t) -c_3 +1/RT\right]}{(c_3 R T-1)^2}
         \tag{FA3}
         \label{eq:EaFA}
\end{equation}
\end{tabular}
&           \begin{tabular}[c]{@{}l@{}}   $c_0$ = -8.518 (1.072) \\ $c_1$ = 0.1266 (0.0191) mol/kcal \\ $c_2$ = -20.99 (5.81) \\ $c_3$ = 0.2985 (0.1026) mol/kcal \\ $\chi_\nu^2$ = 1.66 \\ $0.5\leq n<1$  \end{tabular} \\ \hline
\begin{tabular}[c]{@{}c@{}} Fanning curvilinear \\ (FC) \end{tabular}       &    \begin{tabular}[l]{@{}p{9.5cm}@{}}
\begin{equation}
        f_{FC}(t,T) = c_0 + c_1 \frac{\ln(t)-c_2}{\ln\left(\frac{1}{RT}\right)-c_3}
     \tag{FC1}
    \label{eq:FC}    
\end{equation}
\begin{equation}
         k_{ef}(t,T)_{FC}=\frac{c_1 \exp[(1-n)
f_{FC}(t,T)]}{t \left(\ln \left(\frac{1}{R T}\right)-c_3\right)}
        \tag{FC2}
        \label{eq:kFC}
\end{equation}
\begin{equation}
        E_a^{FC}(t,T)=\frac{R T \left(c_1 c_2 n-c_1 c_2+(c_1-c_1 n) \ln (t)-c_3+\ln \left(\frac{1}{R T}\right)\right)}{\left(c_3-\ln \left(\frac{1}{R T}\right)\right)^2}
         \tag{FC3}
         \label{eq:EaFC}
\end{equation}
\end{tabular}
&  \begin{tabular}[c]{@{}l@{}}   $c_0$ = -9.449 (1.480) \\ $c_1$ = 0.1627 (0.0298) \\ $c_2$ = -24.58 (7.75) \\ $c_3$ = -0.8626 (0.1549) \\ $\chi_\nu^2$ = 1.88 \\ $0.5\leq n<1$ \end{tabular}    \\ \hline
\end{tabular}
}
{\raggedright Notes: 1. For each fission-track annealing model (Eqs. \eqref{eq:PA}, \eqref{eq:PC}, \eqref{eq:CM} \eqref{eq:FA}, \eqref{eq:FC}), the reaction rate constants, $k_{ef}$ and Arrhenius Activation energies were obtained using $g(r) = \ln(1-r)$ and $f_r(r)=(1-r)^n$. 2. Rate constants (Eqs. \eqref{eq:kPA}, \eqref{eq:kPC}, \eqref{eq:kCM} \eqref{eq:kFA}, \eqref{eq:kFC}) were calculated after Eq. \eqref{eq:kef}. 3. Arrhenius activation energies (Eqs. \eqref{eq:EaPA}, \eqref{eq:EaPC}, \eqref{eq:EaCM} \eqref{eq:EaFA}, \eqref{eq:EaFC}) were obtained by the application of Eq.~\eqref{eq:Ea_definition}, and are average values for constant heating experiments.}
\label{tab:synthesis}

\end{table*}

Fission tracks are formed by displaced atoms and vacant sites, in concentrations high enough to change the structure of the mineral in a volume of about 2-10~nm in diameter and around 20 $\mu$m in length. The annealing process is the recombination of defects and vacancies, which also changes the neighbor structure and consequently the recombination rate. This kind of solid-state reaction is described by the conversion rate equation \citep{Vyazovkin2015Book}:

\begin{equation}\label{eq:general conversion rate}
    \frac{\text{d}\alpha}{\text{d}t} = k(T)f_{\alpha}(\alpha).
\end{equation}

The Eq.~\eqref{eq:general conversion rate} relates the rate of conversion of the reactant $\alpha$ with the constant rate and with the reaction function $f_\alpha(\alpha)$. For the fission tracks, $\alpha$ is the concentration of recombined atoms, and $f_\alpha(\alpha)$ describes how the recombination process changes the surrounding structure. The track length can be used as a proxy for the concentration of displaced atoms \citep{rufino2022arrhenius} and:

\begin{equation}\label{eq:fission track conversion rate}
   \alpha=\frac{L_0-L}{L_0}=1-r    
\end{equation}

\noindent and with this change of variable:

\begin{equation}\label{eq:diff rkf}
    \frac{\text{d}r}{\text{d}t}=-k_{ef}(t,T)f_r(r).
\end{equation}

The rate constant has been replaced with an effective rate constant, $k_{ef}(t,T)$, which may depend on time and temperature and is suitable to describe more complex reactions \citep{vyazovkin2016time}. For the reaction function, the reaction-order function has already been shown to produce consistent results mainly for the single activation energy mechanisms of annealing \citep{green1988can, rufino2022arrhenius}: 
%From the Fission-Tracks literature, mainly in the works of \citep{Green1988can}, the idea that annealing is a kinetic process of reaction order $n$ suggests that an ``educated guess'' that holds this property be a reaction function $f_r(r)$ of the type reaction order function:

\begin{equation}\label{eq:reaction order model}
    f_r=(1-r)^n
\end{equation}

\noindent in which $n$ is the reaction order.

Eq.~\eqref{eq:diff rkf} is a differential equation that can be solved by the separation of variables. To define the limits of the integral, consider that at the beginning of the thermal history ($t=0$), the track is unannealed ($r=1$). After a heating duration $t$, the track length has been shortened to $r$. Then:

\begin{equation}\label{eq:rateIntGen}
	\int_r^1 \frac{\text{d}r}{f_r(r)} = \int_0^t -k_{ef}(t,T) \text{d}t.
\end{equation}

Eq.~\eqref{eq:rateIntGen} is the basic equation from which the annealing kinetics can be studied from a physicochemical perspective. Once the reaction function and the rate constant are chosen, the dependence of the reduced fission-track length can be  calculated over any T-t path. Let's start with the known case of constant temperature heating, from which the annealing equations should be obtained. For the single activation energy models, PA, PC, and CM, the rate constants are given by:

\begin{subequations}\label{eq:general k}
    \begin{alignat}{2}
        k_{ef}(T)_{PA} &= A_{1} \exp\left(\frac{-Q_{1}}{R T}\right), \label{eq:general ka}\\
         k_{ef}(T)_{PC} &= A_{2} (RT)^{m}, \label{eq:general kb}\\
         k_{ef}(T)_{CM} &= A_{3} (R T) \exp\left(-\frac{Q_{3}}{R T}\right), \label{eq:general kc}
    \end{alignat}
\end{subequations}

\noindent where $A_i$, $Q_i$, and $m$ are constants. Eq.~\eqref{eq:general ka} is the original Arrhenius equation from which the PA equation is derived. $Q_1$ can be directly identified with the activation energy only in this case. Eq.~\eqref{eq:general kb} generates the PC equation with a temperature-dependent activation energy (Table~\ref{tab:synthesis}, Eq.~\eqref{eq:EaPC}). Eq.~\eqref{eq:general kc} generates the Carlson Model, also with a temperature-dependent activation energy (Table~\ref{tab:synthesis}, Eq.~\eqref{eq:EaCM}). It is the product of Eqs.~\eqref{eq:general ka} and \eqref{eq:general kb}, with $m=1$, and has been proposed soon after the original Arrhenius equation to deal with reactions that deviate from the expected Arrhenius behavior \citep{Kooij1893}. Note that although the activation energies in the Eqs.~\eqref{eq:general kb} and \eqref{eq:general kc} depend on temperature, they still fall into the category of single activation energy processes, meaning that all recombination events at a given temperature have the same activation energy. 

Annealing experiments are isothermal heating procedures. Then, substituting the effective rate constants (Eqs.~\eqref{eq:general k}) into the integral equation (Eq. \ref{eq:rateIntGen}) together with the reaction-order function defined in Eq. \eqref{eq:reaction order model} and solving it considering the temperature as a constant results in: 

\begin{subequations}\label{eq:int solutions}
\begin{alignat}{2}
    \ln(1-r) &= \frac{\ln\left[ A_1(1-n)\right]}{1-n} + \frac{1}{1-n} \ln (t) - \frac{Q_1}{1-n}\frac{1}{RT}, \label{eq:int solution PA}\\ 
     \ln(1-r) &= \frac{\ln[A_2 (1-n)]}{1-n}+\frac{1}{1-n}\ln (t) -\frac{m}{1-n} \ln (\frac{1}{RT}), \label{eq:int solution PC}\\ 
    \ln(1-r) &= \frac{\ln \left[A_3 (1-n)\right]}{1-n}+\frac{1}{1-n}\ln(t) - \frac{Q_3}{1-n}\frac{1}{RT} -\frac{1}{1-n} \ln (\frac{1}{RT}), \label{eq:int solution CM}
\end{alignat}
\end{subequations}

\noindent which are the equations for the PA \eqref{eq:int solution PA}, PC \eqref{eq:int solution PC}, CM \eqref{eq:int solution CM} models with $g(r)=\ln (1-r)$. For the chosen reaction function, the integral only has a real solution if $n < 1$. Comparing the right sides of these equations respectively with Eqs.~\eqref{eq:PA}, \eqref{eq:PC}, and \eqref{eq:CM}, one can find out that the rate constant parameters are related to the fitting parameters of the annealing equations as

\begin{align}
\text{PA}: n&=\frac{c_1-1}{c_1}          &  Q_1 &= -\frac{c_2}{c_1}              &  A_1&=c_1\exp(c_0/c_1) \label{eq:rec PA}\\
\text{PC}: n&=\frac{c_1-1}{c_1}           &  m &=-\frac{c_2}{c_1}              &  A_2&=c_1\exp(c_0/c_1) \label{eq:rec PC}\\
\text{CM}: n&=\frac{c_1-1}{c_1}           &  Q_3 &= -\frac{c_2}{c_1}              &  A_3&=c_1\exp(c_0/c_1)\label{eq:rec CM}
\end{align}

In this way, the rate constants can be expressed in terms of the fitting parameters of the annealing models as shown in Eqs. \eqref{eq:kPA}, \eqref{eq:kPC}, and \eqref{eq:kCM} of Table~\ref{tab:synthesis}. The values for the reaction order $n$ for the three models are $n\approx -4$, in agreement with a similar analysis carried out by \citet{green1988can} for the PA model. Therefore, the parallel models are not compatible with first-order annealing kinetics, meaning that the neighbor structure has a strong influence on the rate of defect recombination during annealing.

There are no obvious expressions for the rate constant for the fanning models. A physicochemical analysis of their trends indicates that multiple concurring processes with different activation energies are occurring during the annealing of the fission tracks \citep{rufino2022arrhenius}, in agreement with previous suggestions \citep{green1988can, tamer2020low}. \citet{rufino2022arrhenius} derived an expression from Eq.~\eqref{eq:diff rkf} to find the effective rate constant from the annealing model:

\begin{equation}\label{eq:kef}
	k_{ef}(t,T) =  -\left.\frac{1}{f_r(r)}\left[ \pdv{g(r)}{r} \right] ^{-1}\pdv{f(t,T)}{t} \right|_{T}
\end{equation}

Eq. \eqref{eq:kef} provides a direct way to calculate this effective reaction rate constant from the model functions that fit the experimental annealing data, $f(t,T)$ and $g(r)$, and from the reaction function $f_r(r)$. The partial derivative in relation to time is taken because the annealing models were designed to describe constant temperature experiments. %Although formulated with a different purpose \citep{rufino2022arrhenius}, Eq.~\eqref{eq:kef} ends up mimicking the PET hypothesis that a length variation only depends on its current value and temperature (see discussion in the Appendix of \citet{duddy1988thermal}).
As a check, before applying Eq.~\eqref{eq:kef} to the fanning models, one can show that Eqs.~\eqref{eq:kPA}, \eqref{eq:kPC}, and \eqref{eq:kCM} are found by the application of Eq.~\eqref{eq:kef} respectively to Eqs.~\eqref{eq:PA}, \eqref{eq:PC}, and \eqref{eq:CM}, with $g(r)=\ln(1-r)$ and $f_r(r)=(1-r)^n$. The same procedure can be applied to Eq.~\eqref{eq:FA} and \eqref{eq:FC} to find the effective reaction rates respectively for the Fanning Arrhenius (Eq.~\eqref{eq:kFA}) and Fanning Curvilinear (Eq.~\eqref{eq:kFC}) models. An alternative way to infer the effective rate constants for the FA and FC models is departing from the hypothesis that the Arrhenius activation energies and, therefore, the rate constants are dependent on the fission-track reduced length. Then, integration of Eq.~\eqref{eq:rateIntGen}, on the isothermal condition, results in

\begin{equation} \label{eq:krIntegral}
    -\int_1^r \frac{\text{d}r}{f_r(r)k_{ef}(r)}=\int_0^t \text{d}t=t.
\end{equation}

It can be shown that the primitive functions that make Eq.~\eqref{eq:krIntegral} true for the FA and FC models, with $f_r(r)=(1-n)^n$ and $g(r)=\ln(1-r)$ are the ones with the effective rate constants given by Eqs.~\eqref{eq:kFA} and \eqref{eq:kFC}. This approach also illustrate how the incorporation of the time in the rate constant and, therefore, in the activation energies for the fanning models are implied from the dependence of the activation energies on the values of $r$.

To obtain the reaction order $n$ for the FA and FC models, the effective reaction rate constants given by Eqs.~\eqref{eq:kFA} and \eqref{eq:kFC} are integrated with Eq.~\eqref{eq:rateIntGen} considering constant temperatures (isothermal experiments):

\begin{subequations}\label{eq:int solutions}
\begin{alignat}{2}
    \int_{1}^{r} \frac{\text{d}r}{(1-r)^n} &= -\int_{0}^{t} \frac{c_1}{\frac{1}{RT}-c_3}\frac{1}{t}\exp\left[-(n-1)\left(c_0+c_1\frac{\ln t - c_2}{\frac{1}{RT}-c_3}\right)\right] \text{d}t \label{eq:int equation FA}\\ 
     \int_{1}^{r} \frac{\text{d}r}{(1-r)^n} &= -\int_{0}^{t} \frac{c_1}{\ln(\frac{1}{RT})-c_3}\frac{1}{t}\exp\left[-(n-1)\left(c_0+c_1\frac{\ln t - c_2}{\ln(\frac{1}{RT})-c_3}\right)\right] \text{d}t\label{eq:int equation FC}
\end{alignat}
\end{subequations}

With the necessary condition of $n<1$, the solution of the integral equation \eqref{eq:int equation FA} is

\begin{equation}
    (1-r)=(-1)^{-1/(n-1)}\exp\left[c_0+c_1\frac{\ln(t)-c_2}{\frac{1}{RT}-c_3}\right]
\end{equation}

As the solution of this equation is to represent the shortening of the fission tracks, $(1-r)$ must be a real value between 0 and 1, which is true only if $-1/(n -1)$ is an even and positive integer value $2j$. Then, the values of $n$ are restricted to

\begin{equation} \label{eq:frac n}
    n = \frac{2j-1}{2j},%\\
    %\text{where $j=1,2,3...$. }
\end{equation}

\noindent where $j=1,2,3,...$. With this condition and $g(r)=\ln(1-r)$, the FA model (Table~\ref{tab:synthesis}, Eq.\eqref{eq:FA}) is recovered. The solutions for Eqs.~\eqref{eq:int equation FA} and \eqref{eq:int equation FC} are similar, differing only on the logarithm of $1/RT$ for the FC model instead of the $1/RT$ for the FA model, which are both constants in this case. The previous analysis holds also for the FC model. The values of $n$ will be fractional for FA and FC ($n=1/2, 3/4, 5/6, 7/8,...$), according to Eq.~\eqref{eq:frac n}. Fractional reaction orders are characteristics of multiple-step reactions or some more complex kinetic mechanism, as it has been explained for the decomposition of acetaldehyde \citep{laidler1965chemical}, a well know example of fractional reaction order in chemistry. However, for the fission tracks, where the displaced atoms and vacant sites take the role of reactants and the deformed track structure is the reaction medium, explanations of the kinetics of a single reactant via a mean-field approximation (MFA) may not be appropriate \citep{cordoba2003fractional}. Thus, for the effective reaction rate constant $k_{ef}$ of the fanning annealing models, mechanistic modeling considering the intermediate steps, i.e., recognizing the reaction order of each mechanism involved in annealing, would be desirable to elucidate the meaning of the fractional reaction order found \citep{koga1992fractional}. The rate constants for FA and FC are to be viewed as effective equations constraining the general behavior of annealing but that does not allow the description of the specifics of the annealing kinetics.   %However, this effective result will prove consistent when taken into account that this fractional reaction order can assume various values within the $1/2\leq n<1$ limit thus expanding, in principle, the analysis to a not properly constrained order kinetics.

In this physicochemical framework of the fission-track annealing, the effective reaction rate constant, $k_{ef}(t,T)$, and the reaction function, $f_r(r)$, are the fundamental building blocks from which the fission-track annealing kinetics can be studied. The application to the constant temperature annealing made it possible to determine the rate constant parameters from the empirically determined parameters of the annealing equations. The calculation of the Arrhenius activation energies ($E_a$) for different models becomes possible through Eq.~\eqref{eq:Ea_definition}. The Arrhenius activation energies of the parallel annealing models (Eq. \ref{eq:EaPA}, \ref{eq:EaPC} and \ref{eq:EaCM}) will be constants with respect to the variable $t$. As for the fanning equations, $E_a$ will vary with time and temperature (Eq.~\ref{eq:EaFA} and \ref{eq:EaFC}). However, the main advantage of this approach is the possibility of calculating the fission-track length reduction over any T-t path using Eq.~\eqref{eq:rateIntGen}, without recurring to the interactive application of the Principle of Equivalent Time.

\subsection{Fission track annealing under variable temperature thermal histories}

Fission-track thermal history inference is based on the Principle of Equivalent Time (PET) \citep{goswami1984quantitative,duddy1988thermal}, which is an interactive method that allows the application of isothermal annealing models to variable temperature T-t paths. It is detailed in Appendix \ref{sec:rPETequations}. In general, a given variable temperature thermal history is divided into finite time intervals $\Delta t_i$, centered at times $t_i$ and temperatures $T_i$. At the time interval in which the population was born, a first reduced length is calculated by applying the annealing equation, using the temperature of the T-t path and the duration of the interval ($T_i,\Delta t_i$). In the next interval, at a different temperature on T-t path, the annealing model is used to find an equivalent time capable of producing the same length shortening of the previous interval but at the new temperature. A new length shortening is then calculated by applying the annealing model to the period of time that is the sum of the equivalent time and the length of the time interval. This procedure is repeated and at any given temperature, $T_i$ on the T-t path, an equivalent time, $\tau_i$, which reproduces the length shortening at the previous interval, $r_{i-1}$, is determined, so that the new length shortening can be calculated as if the track had been at the same constant temperature from the beginning. The reduced length is updated ($r_i$) by calculating it as a result of heating at $T_i$ for the duration $\tau_i+\Delta t_i$. The hypothesis that the annealing kinetics does not depend on the previous thermal history of the track, but only on its current length so that any previous T-t path can be replaced with a constant temperature heating resulting in this length, is the basis of this procedure and defines the Principle of Equivalent Time. This means, in practice, that the track will have no memory of the material conditions of time and temperature of its previous shortenings. The equations for the application of the PET with the PA, PC, CM, FA, and FC models can be found in Table~\ref{tab:PETresults} in Appendix~\ref{sec:rPETequations}.

The physicochemical tool presented in the previous section provides an alternative way to access variable temperature annealing kinetics by solving the integral in the right side of Eq.~\eqref{eq:rateIntGen} over a T-t path. Eq.~\eqref{eq:rateIntGen} is solved as a line integral. A suitable parameterization is:

\begin{equation}\label{eq:Sparametrization}
    s = \begin{cases}
        T=T(u)\\
        t=u
    \end{cases}, \\
    \text{d}s=\sqrt{1+\left(\frac{\text{d}T}{\text{d}u}\right)^2}\text{d}u.
\end{equation}

Implementing the parameterized variables on the right side of the integral equation (Eq. \ref{eq:rateIntGen}),

\begin{equation}\label{eq:intkefparamimplies}
    I = \int_{0}^tk_{ef}\left(t(u),T(u)\right)\frac{\text{d}s}{\sqrt{1+\left(\frac{\text{d}T}{\text{d}u}\right)^2}} \implies I = \int_{0}^tk_{ef}\left(t(u),T(u)\right)\text{d}u.
\end{equation}

Solving the left side of Eq.~\eqref{eq:rateIntGen} for the $f_r(r)$ function given by Eq.~\eqref{eq:reaction order model} and the parameterized integral for the rate constant (Eq.~\eqref{eq:intkefparamimplies}), the reduced length, after the track has experienced the thermal history given by the T-t path, is

\begin{equation}\label{eq:rRCIgeneral}
    r = 1-\left((1-n)\int_{0}^tk_{ef}\left(t(u),T(u)\right)\text{d}u\right)^{1/1-n},
\end{equation}

\noindent in which $n<1$ as usual. At a first glance, the advantage of the Rate Constant path Integral (RCI, Eq.~\eqref{eq:rRCIgeneral}) is that it is a one-shot calculation of the reduced track length. In addition, there was no need to restrict the form of the rate constant function and therefore the annealing mechanism it is related to.

\section{Results and Discussion}
\label{sec:Results}
The RCI Eq.~\eqref{eq:rRCIgeneral} can be applied to calculate the shortening in the reduced length of a single fission-track population submitted to any T-t path. The same calculation can be carried out using the interactive technique based on the Principle of Equivalent Time (PET). To compare the outcomes of the two methods, both calculations will be performed for the parallel (including CM) and fanning models. The case will be made for the linear cooling, $T(t) = T_0 - \dot{T}t$, where $\dot{T}$ is the cooling rate and $T_0$ is the temperature at the time the track was generated. The temperature at the end of the T-t path (present time) was fixed to be 20 $^\circ$C (293.15 K) for this analysis.

\subsection{Parallel models}
\label{sec:resParallel}

\begin{figure}[!h]
\centering
\subfloat[]{\label{fig:rPA}, 
\includegraphics[width=0.32\linewidth]{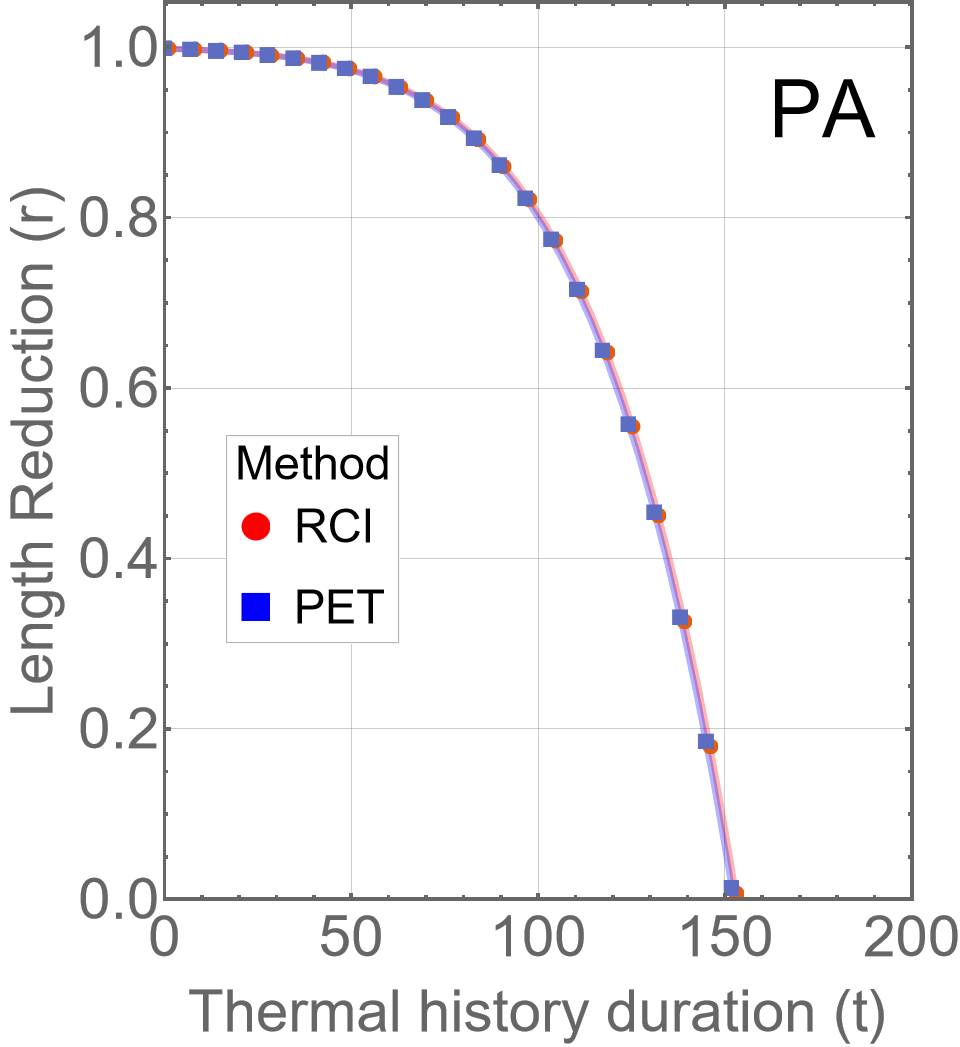}
}
\subfloat[]{\label{fig:rPC}
\includegraphics[width=0.32\linewidth]{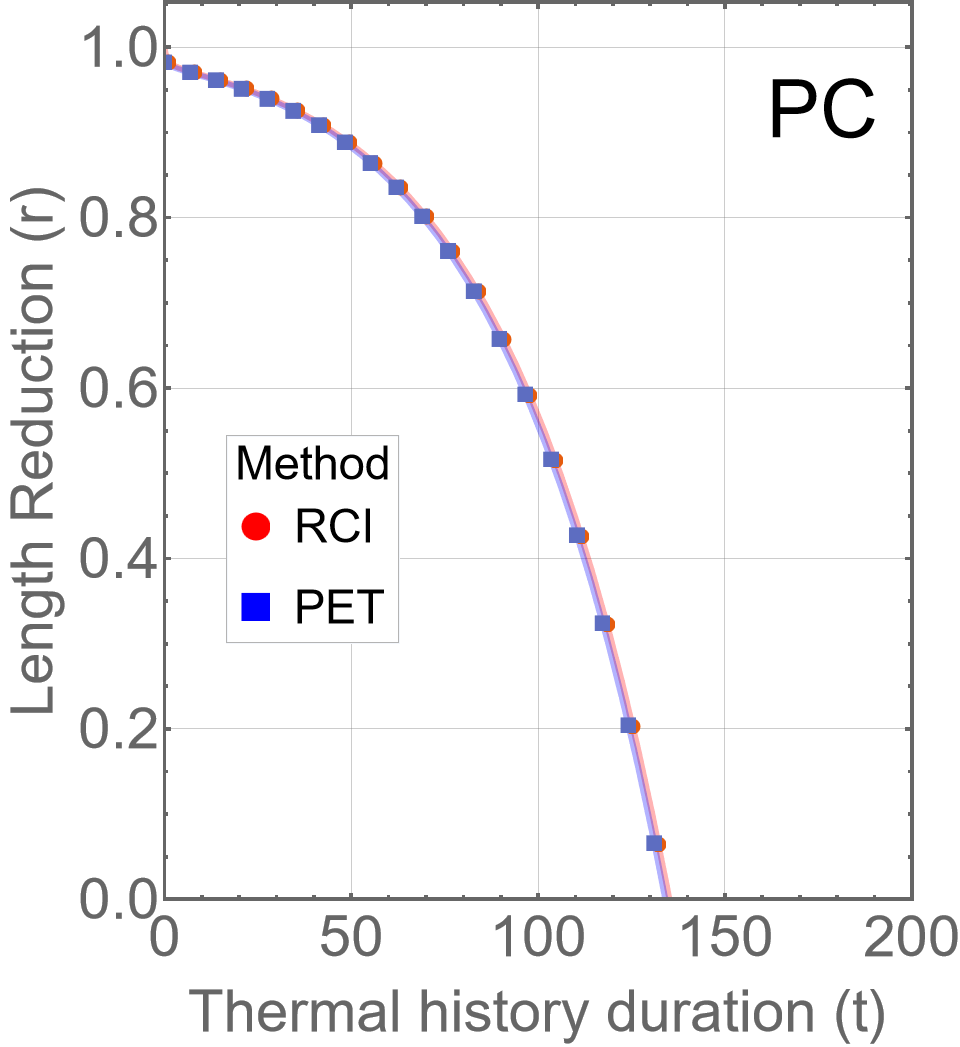}
}
\subfloat[]{\label{fig:rCM}
\includegraphics[width=0.32\linewidth]{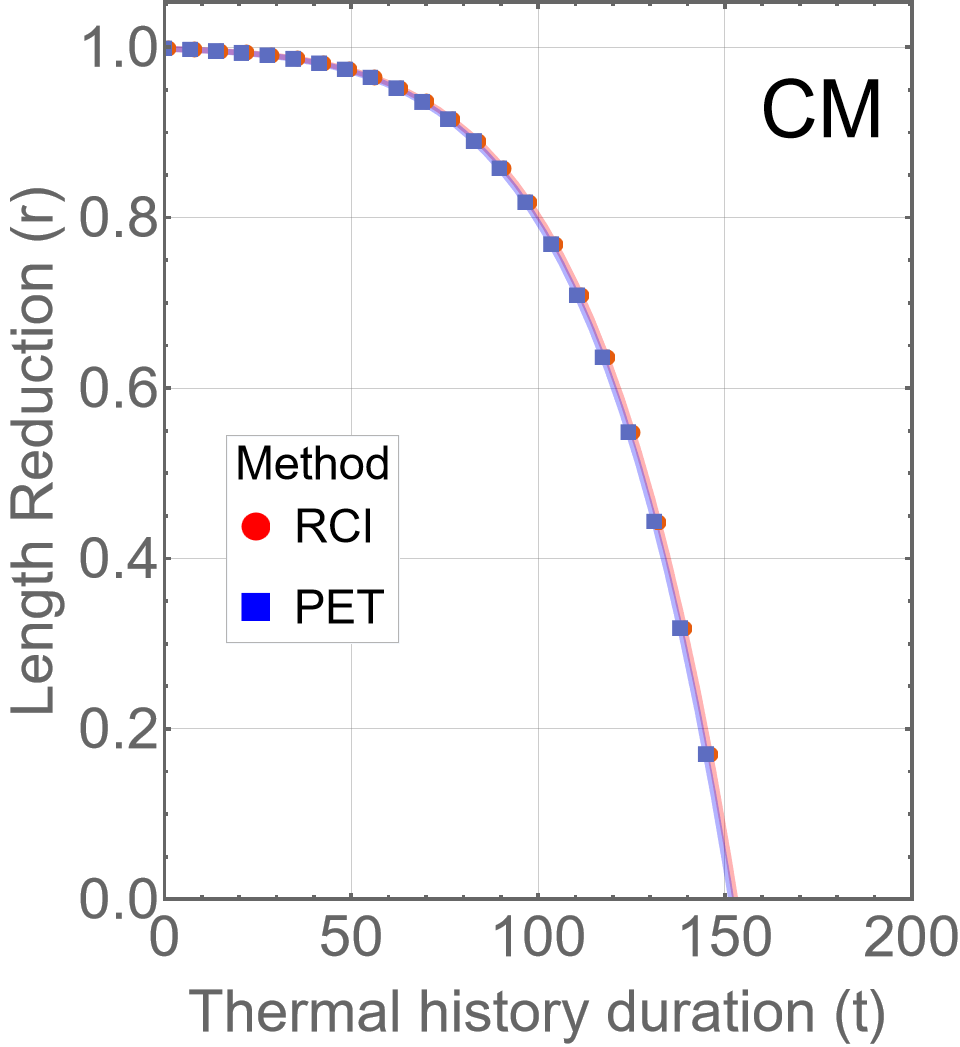}
}
\caption{Values of the reduced track lengths ($r$), after a linear cooling ($\dot{T}=1.0^\circ$C/Ma) starting at time $t$ and ending at the present time at a fixed temperature of 20$^\circ$C, calculated using the Parallel models: (a) Arrhenius, (b) Curvilinear, and (c) Carlson). The points that form the curve in red (circle marks) were calculated by applying the RCI (Eq. \ref{eq:RCI PA}, \ref{eq:RCI PC}, and \ref{eq:RCI CM}). The values calculated using the PET are in blue (square marks). 
}
\label{fig:rParallel}
\end{figure}

To solve the RCI, the effective rate constant functions for the parallel models (Table~\ref{tab:synthesis}, Eqs. \eqref{eq:kPA}, \eqref{eq:kPC}, and \eqref{eq:kCM}) are inserted in Eq.~\eqref{eq:rRCIgeneral} with the variable $T$ replaced with $T(t)=T_0-\dot{T}t$ wherever it appears. The analytical solutions for the reduced length shortening calculated for the PA, PC, and CM are

\begin{subequations}\label{eq:RCI parallels}
\begin{alignat}{2}
    r_{PA} &= 1-\left(\frac{e^{c_0/c_1} \left(c_2 \text{Ei}\left(\frac{c_2}{c_1 R (T_0- \dot{T}t) }\right)-c_2 \text{Ei}\left(\frac{c_2}{c_1 R T_0}\right)+c_1 R \left((\dot{T}  t-T_0) e^{\frac{c_2}{c_1 R (T_0-\dot{T} t)}}+T_0 e^{\frac{c_2}{c_1 R T_0}}\right)\right)}{c_1 \dot{T}  R}\right)^{c_1},\label{eq:RCI PA}\\ 
    r_{PC} &= 1-\left(\frac{c_1 e^{c_0/c_1} \left((\dot{T}  t-T_0) (R (T_0-\dot{T}  t))^{-\frac{c_2}{c_1}}+T_0 (R T_0)^{-\frac{c_2}{c_1}}\right)}{c_1 \dot{T} -c_2 \dot{T} }\right)^{c_1},\label{eq:RCI PC}\\
\begin{split}
    r_{CM} =  1-2^{-c_1}e^{c_0}\frac{ c_1^{-2 c_1}}{(\dot{T}  R)^{c_1}}\left[c_1 R \left(T_0 e^{\frac{c_2}{c_1 R T_0}} (c_1 R T_0+c_2)-(T_0-\dot{T}  t) e^{\frac{c_2}{c_1 R T_0-c_1 \dot{T}  R t}} (c_1 R (T_0-\dot{T}  t)+c_2)\right)\right.\\
 +c_2^2 \left.\left(\text{Ei}\left(\frac{c_2}{c_1 R T_0-c_1 R t \dot{T} }\right)-\text{Ei}\left(\frac{c_2}{c_1 R T_0}\right)\right)\right]^{c_1},\label{eq:RCI CM}
\end{split}
\end{alignat}
\end{subequations}

\noindent where Ei is the exponential integral function. Eqs. \eqref{eq:RCI PA} - \eqref{eq:RCI CM} give the resulting reduced length $r$ for the parallel models, as functions of the three variables that characterize the thermal history: the duration of the T-t path ($t$), the cooling rate ($\dot{T}$), and the temperature at the time when the track was born ($T_0$). The parameters $c_i$ are given in the last column of Table~\ref{tab:synthesis}. The values of $r$ for the cooling path with the cooling rate $\dot{T}=1.0 ^\circ$C/Ma calculated with the three parallel models are presented in Fig.~\ref{fig:rParallel}. For each point, the value of $r$ is the length reduction after a linear cooling duration $t$ and measured in the present. Values of $r=0$ mean that the tracks have been erased before the present. Values obtained by the RCI solutions (Eqs.~\eqref{eq:RCI PA} - \eqref{eq:RCI CM}) are represented as red curves marked with red circles and the values calculated using the PET are represented as blue curves marked with blue squares. RCI and PET calculations produce very close values of $r$ for the three parallel models (Figs.~\ref{fig:rPA}, \ref{fig:rPC}, \ref{fig:rCM}).

The temperature indexes Closure Temperature ($T_C$) and Total Annealing Temperature ($T_A$) were also calculated for the three parallel models, applying both methods of calculation, for cooling T-t paths with cooling rates of 1, 10, and 100 $^\circ$C/Ma. $T_C$ is, for a monotonic cooling thermal history, the temperature at the apparent sample age \citep{dodson1973closure}. $T_A$ is the age of the oldest track that has not been erased and can be counted in the sample \citep{issler1996TA}. Details for the method of calculation of the index temperatures can be found in \citet{guedes2013improved}. $T_C$ and $T_A$ are meaningful quantities that allow quantifying the impact of using the RCI instead of the interactive PET calculation. The uncertainties in $T_C$ and in $T_A$ were estimated by simple error propagation of apparent ($T_C$) or retention ($T_A$) ages and present time temperature. Results are shown in Table~\ref{tab:thermal_indexes}. Setting the PET results as the reference values, given that PET is the method established in the literature, a relative error analysis can be carried out to verify the internal consistency between PET and RCI calculations. The relative error between PA PET and PA RCI is on average 0.69$\%$ for $T_C$ and 1.19$\%$ $T_A$. The same trend is found for calculations of $T_C$ and $T_A$ with PC and CM:  0.66$\%$ and 0.58$\%$ (PC) and 0.7$\%$ and 1.19$\%$ (CM). All errors are well inside the estimated uncertainties for temperature index values and are most probably artifacts of the PET numerical calculation.

%table for T0 = 20
\begin{table}[]
	\caption{Comparison of thermal indexes for the models presented in this work, calculated using the PET and the RCI}
	\resizebox{\textwidth}{!}{
		\begin{tabular}{lcccccc}
			\hline 
			Principle of Equivalent Time (PET) thermal indexes \\
			Fit                                      & $T_C$1     & $T_C$10    & $T_C$100   & $T_A$1     & $T_A$10    & $T_A$100   \\ \hline
			PA                                        & 130(7) & 144(7) & 158(8) & 153(8) & 168(8) & 184(9) \\
			PC                                        & 105(6)  & 121(6)  & 139(7)  & 132(8) & 151(8) & 171(9) \\
			CM                                        & 130(7)  & 143(7)  & 157(8)  & 153(8) & 168(8) & 184(9) \\
			FA                                        & 134(7) & 146(7) & 160(8) & 163(8) & 176(8) & 191(10) \\
			FC                                        & 111(6) & 126(6) & 143(7) & 143(7) & 160(8) & 179(9) \\ \hline
			Rate constant integral on a path (RCI) thermal indexes \\
			Fit                                      & $T_C$1     & $T_C$10    & $T_C$100   & $T_A$1     & $T_A$10    & $T_A$100   \\ \hline
			PA                                        & 131(7) & 145(7) & 159(8) & 155(8) & 170(8) & 186(9) \\
			PC                                        & 106(6)  & 122(6)  & 140(7)  & 133(8) & 152(7) & 172(9) \\
			CM                                        & 131(7)  & 144(7)  & 158(8)  & 155(8) & 170(8) & 186(9) \\
			FA                                        & 125(7) & 137(7) & 150(7) & 153(9) & 166(8) & 181(10) \\
			FC                                        & 100(6) & 115(6) & 130(6) & 130(8) & 148(8) & 165(8) \\ \hline
		\end{tabular}
		\label{tab:thermal_indexes}}
	\raggedright Notes: 1. Temperatures in $^\circ$C. 2. The standard errors are given in parentheses.  3. $T_C$: Closure Temperature. 4. $T_A$: Total Annealing Temperature. 5. The numbers to the right of $T_C$ and $T_A$ are the cooling rates in $^\circ$C/Ma.
\end{table}

The PET was formulated under the hypothesis that the annealing of fission tracks is a single activation energy process \citet{duddy1988thermal}. The internal consistency between PET and RCI values of $r$, $T_C$, and $T_A$ calculated with the parallel models is a check for the robustness of the physicochemical approach to deal with variable temperature thermal histories. It is to be noted that not only the PA model, in which the activation energy is temperature-independent (Table~\ref{tab:synthesis}, Eq.~\eqref{eq:EaPA}), but also the PC model, in which the activation energy is temperature-dependent (Table~\ref{tab:synthesis}, Eqs.~\eqref{eq:EaFC}), show such internal consistency. The same agreement is observed for CM. The CM activation energy may vary with temperature but, with the parameters shown in Table~\ref{tab:synthesis}, its Arrhenius activation energy is approximately constant since the value of $c_2/c_1$ (54.9 kcal/mol) is much higher than typical values of $RT$ (< 1.0 kcal/mol). Although the activation energies may vary with temperature, these models imply that at any given temperature, the recombination events are taking place with the same activation energy. This is a sufficient condition for the applicability of the PET.

\subsection{Fanning models}
\label{sec:resFanning}

\begin{figure}[!h]
\centering
    \subfloat[]{\label{fig:rFA}
    \includegraphics[width=0.4\linewidth]{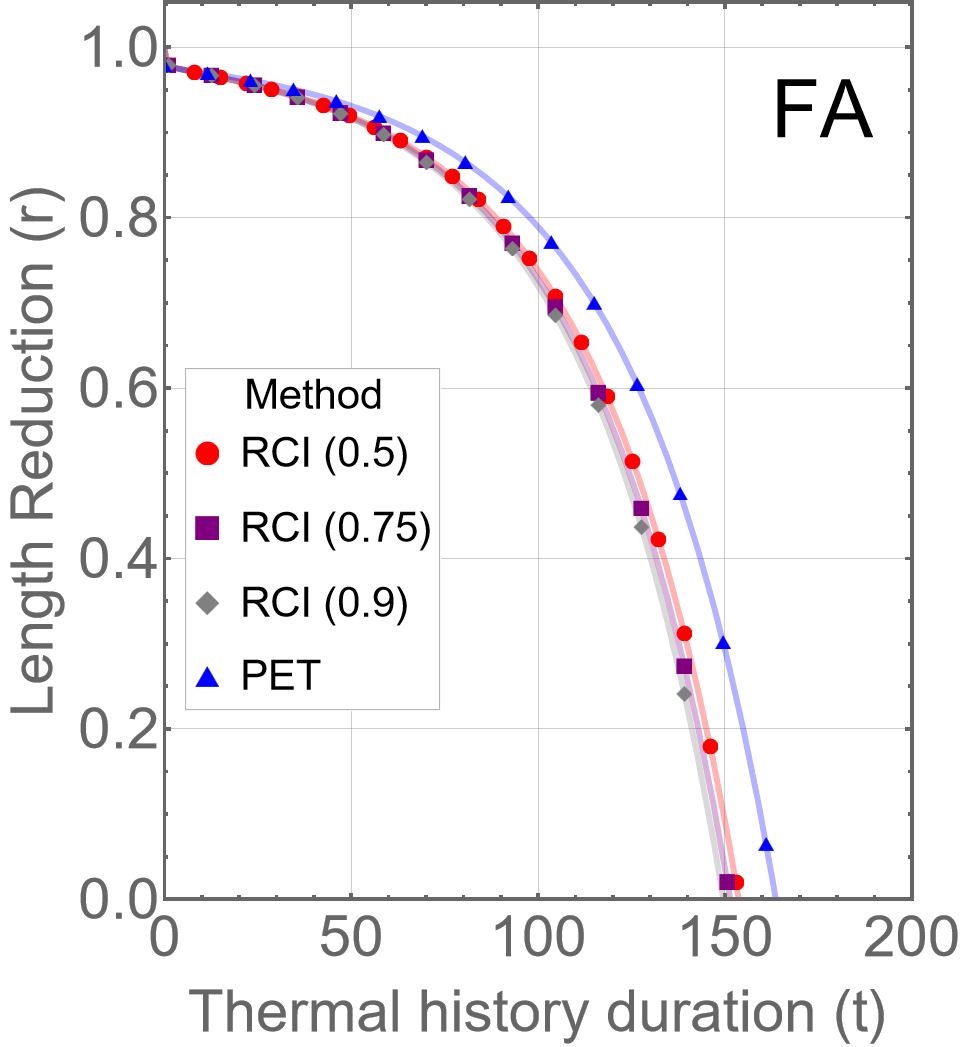}
}
    \subfloat[]{\label{fig:rFC}
    \includegraphics[width=0.4\linewidth]{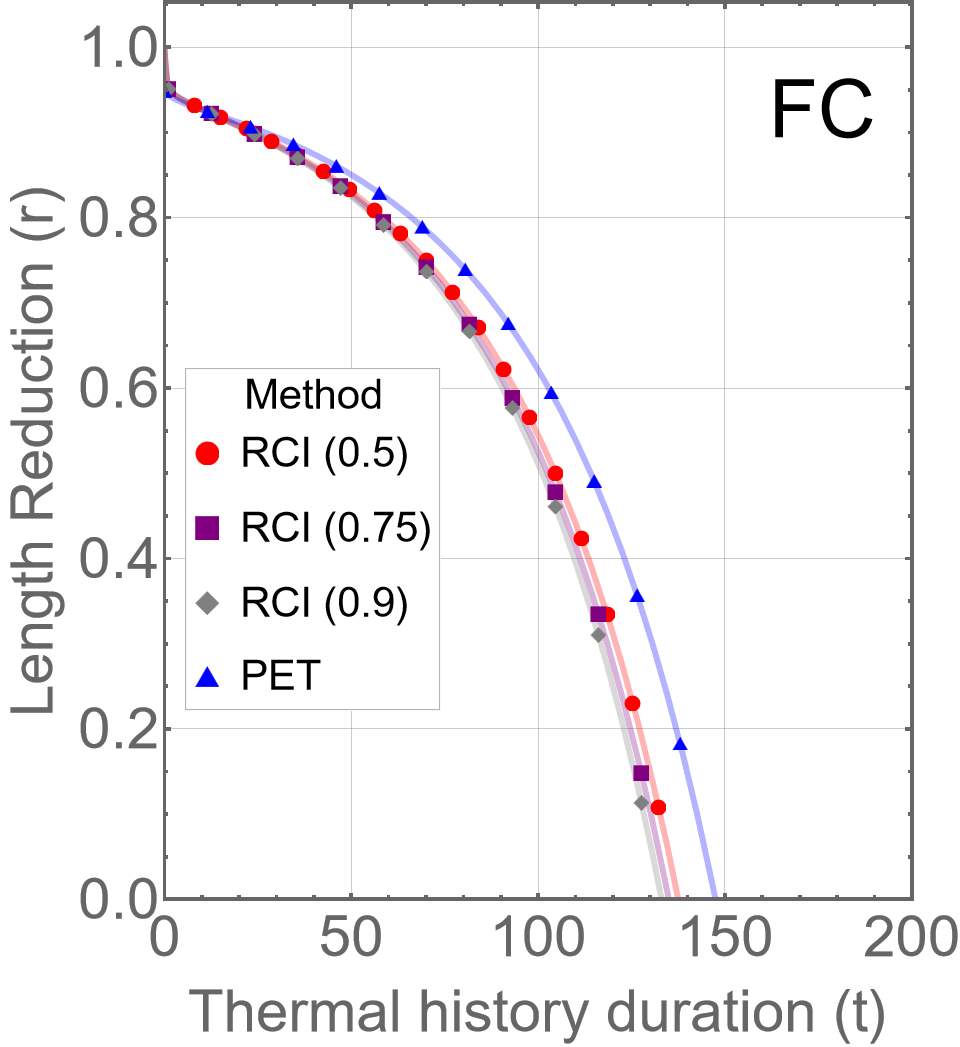}
}
\caption{Values of the reduced track lengths ($r$), after a linear cooling ($\dot{T}=1.0^\circ$C/Ma) starting at time $t$ and ending at the present time at a fixed temperature of 20$^\circ$C, calculated using the Fanning models: (a) Arrhenius and (b) Curvilinear. The points that form the curves in red (circle marks), purple (square marks), and light purple (diamond marks) were calculated by applying the RCI, respectively with $n=0.5$, $n=0.75$, and $n=0.9$. The values calculated using the PET are in blue.
}
\label{fig:rFAFC}
\end{figure}

The values of $r$, $T_C$, and $T_A$ for the cooling T-t path were also calculated for the fanning models, using both the interactive PET and RCI methods. However, the RCI (Eq.~\eqref{eq:rRCIgeneral}) could not be solved analytically for the FA and FC rate constants (Table~\ref{tab:synthesis}, Eqs.~\eqref{eq:kFA} and \eqref{eq:kFC}). The integrals were then solved numerically with the Wolfram Mathematica software \citep{Mathematica}. For validation, the integrals for the parallel models were also solved numerically resulting in exactly the same values obtained with the analytical solutions (Eqs.~\eqref{eq:RCI PA}-\eqref{eq:RCI CM}). Another feature to be considered is that there are certain fractional values allowed for the reaction order $n$, given by Eq.~\eqref{eq:frac n}. The analysis will be limited to $n= 0.5$, $n=0.75$ and $n=0.9$. The numerical method breaks down when $n>0.95$ although its mathematical upper bound is $n<1$. The reduced length calculation results are shown in Fig.~\ref{fig:rFAFC}. Values calculated with the PET are shown in blue, with triangle marks, while values found by solving the RCI are shown, in red ($n=0.5$), purple ($n=0.75$), and light purple ($n=0.9$), respectively with circle, square, and diamond marks. RCI $r$ curves are very close to each other but depart from the $r$ values calculated with the PET. Significant differences between RCI and PET $r$ values are observed for the FC (Fig.~\ref{fig:rFC}) and for the FA (Fig.~\ref{fig:rFA}) models.

The values of $T_C$ and $T_A$ for the fanning models, calculated using the PET and the RCI ($n=0.5$) are in Table~\ref{tab:thermal_indexes}. For the FA model, the mean relative errors between the PET and the RCI $T_C$ and $T_A$ calculations are respectively 6.25$\%$ and 5.68$\%$. For the FC model, the same comparisons result in still more significant differences: 9.57$\%$ ($T_C$) and 8.48$\%$ ($T_A$). The deviations between the values calculated using the PET and the RCI are much more significant than the ones found for the parallel model calculations. One major issue is that the fanning models do not fulfill the single activation energy hypothesis on which the PET is founded. The fanning models emerge from multiple concurring processes with different activation energies \citep{tamer2020low, rufino2022arrhenius}. The effective Arrhenius activation energies incorporate a time dependence (Table~\ref{tab:synthesis}, Eqs.~\eqref{eq:EaFA} and \eqref{eq:EaFC}) that is the consequence of their dependence on the current reduced fission-track length (different slopes of the isoretention curves in the pseudo-Arrhenius plot). On the other hand, the rate constant integral (Eq.~\eqref{eq:rRCIgeneral}) was obtained in a physicochemical framework developed to deal with chemical reactions that did not fit the single activation energy Arrhenius law \citep{Vyazovkin2015Book}. It is by design suitable for complex activation energy systems like the ones pictured by the fanning models. Note also that the presented figures are particular of the fitting parameters in Table~\ref{tab:synthesis}. A different set of parameters would result in different values without changing the conclusion that RCI and PET predictions deviate from each other.

\subsection{Implications for the thermal history modeling}
\label{sec:implications}

The fanning models, especially the Fanning Curvilinear, have been shown to produce better fits to laboratory data and better geological extrapolation of annealing effects \citep{ketcham2007improved, guedes2013improved}. However, the application of the FC along with the PET is an approximation. Compared with the RCI formulation, it underestimates the annealing effect in about 10$\%$, i.e., it predicts that higher temperatures are necessary for the same length shortening as calculated with the RCI for the tested cooling histories. In the context of the inverse problem of inferring T-t paths from the FT age and track length distribution of a mineral sample, it implies the requirement of a longer residence time in the partial annealing zone. For instance, compare, in Table~\ref{tab:thermal_indexes}, the FC $T_A$ calculated with RCI for a cooling rate of 10$^\circ$C/Ma (148$^\circ$C) with the FC $T_A$ calculated with PET for a cooling rate of 1$^\circ$C/Ma (143$^\circ$C). The same analysis applies to the FA model with less significant relative error figures (about 6$\%$).

\begin{figure}[ht]
\centering
%\subfloat{
    \includegraphics[width=0.4\linewidth]{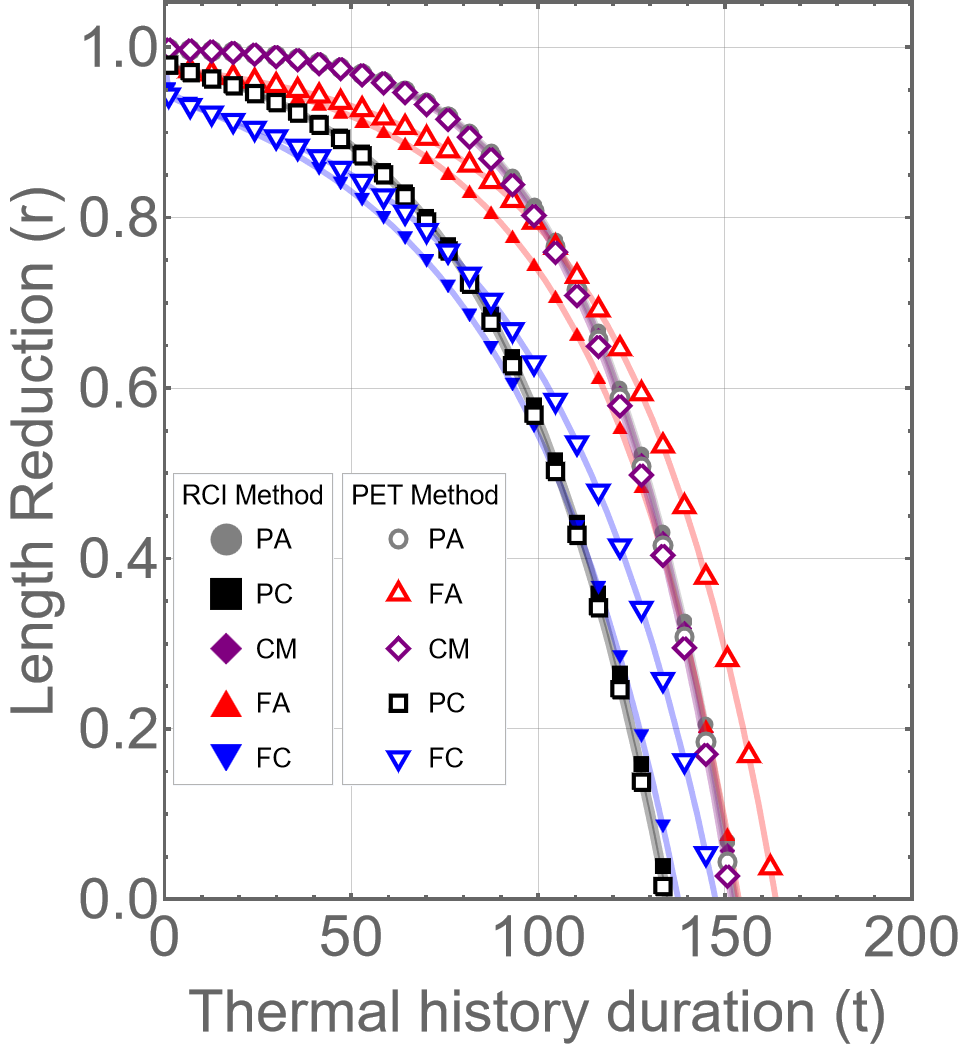}
%}
%\subfloat{
%\vspace{20mm}
%\centering
%\begin{tabular}[]{lCll}
%        &\textbf{Mean Relative Error (MRE)}\\
%        \hline
%        Comparison & Method & $T_C$ & TA   \\ 
%        \hline
%        FA-PA & ETP-ETP & 1.39$\%$ & 4.76$\%$ \\
%        PA-PA & ETP-RCI & 0.69$\%$ & 1.19$\%$  \\
%       PC-PC & ETP-RCI & 0.83$\%$ & 0.66$\%$  \\
%        CM-CM & ETP-RCI & 0.70$\%$ & 1.19$\%$  \\
%        FA-FA & ETP-RCI & 6.25$\%$ & 5.68$\%$  \\
%        FC-FC & ETP-RCI & 10.32$\%$ & 10.00$\%$  \\
%        \hline
%\end{tabular}
%}
\caption{Values of the reduced track lengths ($r$), after a linear cooling ($\dot{T}=1.0^\circ$C/Ma) starting at time $t$ and ending at the present time at a fixed temperature of 20$^\circ$C, calculated for the parallel and fanning models. Calculations using RCI are shown as solid geometric forms, while calculations using PET are represented by empty geometric forms.
}
\label{fig:rAll}
\end{figure}

The Parallel models (PA, PC, and CM), which can be safely applied along with the PET, have long been ruled out for FTT studies \citep{laslett1987thermal, guedes2013improved, Ketcham1999, ketcham2007improved, Ketcham2019}. \citet{duddy1988thermal} had argued that the FA deviated only slightly from the PA model and applied it along with the PET. The isoretention curves for the two models follow approximately the same trends (Fig.~\ref{fig:pseudoArrheniusLinear}). The same behavior is observed for the curvilinear models (Fig.~\ref{fig:pseudoArrheniusCurvilinear}). FC and PC isoretention curves bend together towards lower temperatures.  Their argument can be better appreciated in Fig.~\ref{fig:rAll}. All the PET and RCI predictions for reduced lengths after the track underwent the cooling history are gathered in the same plot. Note that the linear models (PA and FA) and the approximately linear CM form a cluster, while the curvilinear models (PC and FC) form a separate set. The predictions with the fanning models and PET are closer to the predictions of the parallel models for track populations born when the sample passed through intermediate temperatures (partial annealing zone), which results in closer $T_C$ values (compare values in Table~\ref{tab:thermal_indexes}). For populations born at higher temperatures, the fanning-PET predictions depart from the parallel model ones, resulting in a more significant difference between calculated $T_A$ values. Calculations with RCI approximate fanning and parallel model predictions for populations born at higher temperatures. Within this approximation, it could be possible to engineer fanning model parameters to make the model even closer to a parallel model.

\section{Concluding remarks}
\label{Conclusion}

Departing from Eq.~\eqref{eq:rateIntGen}, a physicochemical framework was built to deal with the effects of annealing in variable temperature T-t paths. The basic building blocks are the reaction function $f_r(r)$ and the effective rate constant, $k_{ef}(t,T)$. The parallel models (PA, PC, and CM) were shown to be consistent with the single activation energy rate constants given by Eqs.~\eqref{eq:general ka}-\eqref{eq:general kc} and with the reaction-order function (Eq.~\eqref{eq:reaction order model}). The fanning models (FA and FC) are the representation of multiple concurrent recombination processes with different activation energies \citep{rufino2022arrhenius}. The $k_{ef}(t,T)$ functions were built (Eq.~\eqref{eq:kef}) to be consistent with the reaction-order function (Eq.\eqref{eq:reaction order model}). Obtaining FA and FC rate constants from first principle, i.e., a composition of rate constants for individual processes, and validating them experimentally is still an open issue that has to be dealt with. The Eq.~\eqref{eq:rRCIgeneral} is the line integral related to Eq.~\eqref{eq:rateIntGen} from which length shortening due to cooling T-t paths can be directly calculated, independently of whether the rate constant represents a single or multiple activation energy mechanism.

The Principle of Equivalent Time, on the other hand, is only valid for single activation energy equations, which, for the fission-track system, are the parallel models. In these cases, the RCI-based calculations are in agreement with the PET ones (Fig.~\ref{fig:rParallel}), indicating the robustness of RCI formulation. For the fanning models, the use of the PET has long been recognized as an approximation \citep{duddy1988thermal}. Deviations have indeed been observed between RCI and PET-based calculations (Fig.~\ref{fig:rFAFC}). Compared to the application of RCI, the PET calculation underestimates annealing effects in variable temperature T-t paths (Table~\ref{tab:thermal_indexes}).

The PET along with FA or FC models is the calculation method used to infer most published thermal histories. This procedure introduces a systematic deviation that should be considered in the geological interpretation of the thermal history modeling. Alternatively, the rate constant integral (Eq.~\eqref{eq:rRCIgeneral}) could be considered to substitute the PET in inversion thermal history codes \citep{ketcham2005software, gallager2012software}. Computationally, solving an integral, even numerically, is a routine faster than the interactive steps necessary to apply the PET. More importantly, if the rate constants are representative of the track annealing kinetics, this framework results, in principle, in more accurate predictions of the annealing effects in samples submitted to variable temperature thermal histories.

\section*{Appendix}
\appendix
\label{Appendixes}
\section{Equations for the application of the Principle of Equivalent Time}
\label{sec:rPETequations}

\setcounter{table}{0}
\renewcommand{\thetable}{A\arabic{table}}
\begin{table}[]%\label{tab:PETresults}

\caption{Equations for the application of the Principle of Equivalent Time}
\resizebox{\textwidth}{!}{
\setlength\abovedisplayskip{1pt}
\setlength\belowdisplayskip{1pt}
\begin{tabular}{lcl}
\hline
FT models &  Equations    & Parameters (standard error)  \\ \hline
\begin{tabular}[c]{@{}c@{}} Parallel Arrhenius \\ (PA) \end{tabular}        & \begin{tabular}[l]{@{}p{9.5cm}@{}}
%\begin{equation}   
%f_{PA}(t,T)= c_0 + c_1 \ln(t) + \frac{c_2}{R T}  
%\tag{PA1} \label{eq:PA}  
%\end{equation}
\begin{equation}
          \ln \left(1-r_{j-1}\right)=1-\exp \left[f_{PA}\left(\text{$\Delta $t}_j+\tau _{r_{j-1}},T_j\right)\right]
         \tag{PA4}
         \label{eq:PTErPA}
\end{equation}
\begin{equation}
          \ln(\tau_{r_{j-1}})=\frac{\ln \left(1-r_{j-1}\right)}{c_1}-\frac{c_2}{c_1 R T_j}-\frac{c_0}{c_1}
          \tag{PA5}
          \label{eq:PTEtauPA}
\end{equation}
\end{tabular}
    & \begin{tabular}[c]{@{}l@{}} $c_0$ = 5.631 (0.220) \\ $c_1$ = 0.1865 (0.0066) \\ $c_2$ = -10.46 (0.31) kcal/mol\\ $\chi_\nu^2$ = 2.65 \end{tabular} \\ \hline
\begin{tabular}[c]{@{}c@{}} Parallel curvilinear \\ (PC) \end{tabular}       &  \begin{tabular}[l]{@{}p{9.5cm}@{}}
%\begin{equation}
%        f_{PC}(t,T) = c_0 + c_1 \ln(t) +c_2\ln\left(\frac{1}{R T} \right)
%    \tag{PC1}
%    \label{eq:PC}
%\end{equation}
\begin{equation}
         \ln \left(1-r_{j-1}\right)=1-\exp \left[f_{PC}\left(\text{$\Delta $t}_j+\tau _{r_{j-1}},T_j\right)\right]
         \tag{PC4}
         \label{eq:PTErPC}
\end{equation}
\begin{equation}
         \ln(\tau_{r_{j-1}})=\frac{\ln \left(1-r_{j-1}\right)}{c_1}
         -\frac{c_2}{c_1}\ln \left(\frac{1}{R T_j}\right)-\frac{c_0}{c_1}
         \tag{PC5}
         \label{eq:PTEtauPC}
\end{equation}
\end{tabular}
& \begin{tabular}[c]{@{}l@{}} $c_0$ = -4.910 (0.096) \\ $c_1$ = 0.1944 (0.0060) \\ $c_2$ = -9.610 (0.244) \\ $\chi_\nu^2$ = 2.12 \end{tabular}  \\ \hline
\begin{tabular}[c]{@{}c@{}} Carlson Model \\ (CM) \end{tabular}       &  \begin{tabular}[l]{@{}p{9.5cm}@{}}
%\begin{equation}
%        f_{PPC}(t,T) = c_0 + c_1\left(c_3+\ln(t) -\frac{c_2}{RT}+\ln(R T)\right)
 %       \tag{CM1}
 %       \label{eq:CM}
%\end{equation}
\begin{equation}
         \ln \left(1-r_{j-1}\right)=1-\exp \left[f_{CM}\left(\text{$\Delta $t}_j+\tau _{r_{j-1}},T_j\right)\right]
         \tag{CM4}
         \label{eq:PTErCM}
\end{equation}
\begin{equation}
          \ln(\tau_{r_{j-1}})=\frac{\ln \left(1-r_{j-1}\right)}{c_1}-\frac{c_2}{c_1 R T_j}-\frac{c_0}{c_1}-\ln \left(R T_j\right)
         \tag{CM5}
         \label{eq:PTEtauCM}
\end{equation}

\end{tabular}
& \begin{tabular}[c]{@{}l@{}} $c_0$ = 5.426 (0.2155) \\ $c_1$ = 0.1867 (0.0066) \\ $c_2$ = -10.25 (0.2994) kcal/mol \\  $\chi_\nu^2$ = 2.63 \end{tabular}  \\ \hline
\begin{tabular}[c]{@{}c@{}} Fanning Arrhenius \\ (FA) \end{tabular}       &  \begin{tabular}[l]{@{}p{9.5cm}@{}}
%\begin{equation}
%        f_{FA}(t,T) = c_0 + c_1 \frac{\ln(t)-c_2}{\frac{1}{RT}-c_3}
%        \tag{FA1}
%        \label{eq:FA}
%\end{equation}
\begin{equation}
         \ln \left(1-r_{j-1}\right)=1-\exp \left[f_{FA}\left(\text{$\Delta $t}_j+\tau _{r_{j-1}},T_j\right)\right]
         \tag{FA4}
         \label{eq:PTErFA}
\end{equation}
\begin{equation}
         \ln(\tau_{r_{j-1}})=\frac{\left[\ln \left(1-r_{j-1}\right)-c_0\right] \left[\frac{1}{R T_j}-c_3\right]}{c_1}+c_2
         \tag{FA5}
         \label{eq:PTEtauFA}
\end{equation}
\end{tabular}
&           \begin{tabular}[c]{@{}l@{}}   $c_0$ = -8.518 (1.072) \\ $c_1$ = 0.1266 (0.0191) mol/kcal \\ $c_2$ = -20.99 (5.81) \\ $c_3$ = 0.2985 (0.1026) mol/kcal \\ $\chi_\nu^2$ = 1.66 \end{tabular} \\ \hline
\begin{tabular}[c]{@{}c@{}} Fanning curvilinear \\ (FC) \end{tabular}       &    \begin{tabular}[l]{@{}p{9.5cm}@{}}
%\begin{equation}
%        f_{FC}(t,T) = c_0 + c_1 \frac{\ln(t)-c_2}{\ln\left(\frac{1}{RT}\right)-c_3}
%     \tag{FC1}
%    \label{eq:FC}    
%\end{equation}
\begin{equation}
        \ln \left(1-r_{j-1}\right)=1-\exp \left[f_{FC}\left(\text{$\Delta $t}_j+\tau _{r_{j-1}},T_j\right)\right]
        \tag{FC4}
        \label{eq:PTErFC}
\end{equation}
\begin{equation}
          \ln(\tau_{r_{j-1}})=\frac{\left[\ln \left(1-r_{j-1}\right)-c_0\right] \left[\ln(\frac{1}{R T_j})-c_3\right]}{c_1}+c_2
          \tag{FC5}
          \label{eq:PTEtauFC}
\end{equation}

\end{tabular}
&  \begin{tabular}[c]{@{}l@{}}   $c_0$ = -9.449 (1.480) \\ $c_1$ = 0.1627 (0.0298) \\ $c_2$ = -24.58 (7.75) \\ $c_3$ = -0.8626 (0.1549) \\ $\chi_\nu^2$ = 1.88  \end{tabular}    \\ \hline
\end{tabular}
}
{\raggedright Notes: 1. For each fission-track annealing model (Eqs. \eqref{eq:PA}, \eqref{eq:PC}, \eqref{eq:FA}, \eqref{eq:FC}), the length reduction $r$ (Eqs. \eqref{eq:PTErPA}, \eqref{eq:PTErPC}, \eqref{eq:PTErFA}, \eqref{eq:PTErFC}) and equivalent time $\tau$ (Eqs. \eqref{eq:PTEtauPA}, \eqref{eq:PTEtauPC}, \eqref{eq:PTEtauFA}, \eqref{eq:PTEtauFC}) were obtained using $g(r) = \ln(1-r)$.} %and $f_r(r)=(1-r)^n$
\label{tab:PETresults}

\end{table}

The proposed method to deal with annealing in thermal histories with variable temperatures is also based on the Arrhenius equation and was first proposed by \cite{ goswami1984quantitative}. The Principle of Equivalent Time (PET), on which this method is founded, states that the annealing rate of a track does not depend on its previous thermal history, but only on its current length.

The procedure to solve this problem is to infer the magnitude of annealing recursively, dividing the thermal history into appropriate intervals $\Delta t_i$ and starting from a given value of $r$. For each step, annealing is carried out at constant temperature. In the first step, centered at $(t_1, T_1)$. The procedure will be shown for the Parallel Arrhenius model. The reduced fission-track length is calculated using Eq.~\eqref{eq:PA}, along with $g(r) = \ln(1-r)$:

\begin{equation}
    r_1 = 1-(\Delta t_1)^{c_1}\exp(c_0+\frac{c_2}{R T_1})
\end{equation}

For the next step, the calculation of $r_2$ incorporates the principle in the form that this new shortening is postulated to be independent of the previous one. Thus, one can find a length of time, $\tau_{r_1}$, that will yield the length $r_1$ but for heating at the temperature of the second interval, $T_2$:

\begin{equation}
    \tau_{r_1} = (1-r_1)^{-1/c_1}\exp(-\frac{1}{R T_2}\frac{c_2}{c_1}+\frac{c_0}{c_1}).
\end{equation}

The value of $r_2$ is then found by the application of the annealing model to the interval $\tau_{r_1} + \Delta t_2$:

\begin{equation}
    r_2 = 1-(\tau_{r_1}+ \Delta t_2)^{c_1}\exp(c_0+\frac{c_2}{R T_2}).
\end{equation}

This procedure is interactively repeated for the entire T-t path. The last value of $r$ will be the reduced length of the population born in the first interval after experiencing the entire thermal history. For each interval $j$, the formulas above are:

\begin{subequations}
\begin{eqnarray}
     r_{j}&= 1-(\tau_{r_{j-1}}+ \Delta t_{j})^{c_1}\exp(c_0+\frac{c_2}{R T_{j}})\\
    \tau_{r_{j-1}}&= (1-r_{j-1})^{-1/c_1}\exp(-\frac{1}{R T_{j}}\frac{c_2}{c_1}+\frac{c_0}{c_1})
\end{eqnarray}
\end{subequations}

The formulas for the PA, PC, CM, FA, and FC models are presented in Table~\ref{tab:PETresults}.

\section*{Acknowledgements}
This work has been funded by grant 308192/2019-2 by the National Council for Scientific and Technological Development (Brazil).

\bibliographystyle{cas-model2-names}

\bibliography{cas-refs}

\begin{thebibliography}{34}
\expandafter\ifx\csname natexlab\endcsname\relax\def\natexlab#1{#1}\fi
\providecommand{\url}[1]{\texttt{#1}}
\providecommand{\href}[2]{#2}
\providecommand{\path}[1]{#1}
\providecommand{\DOIprefix}{doi:}
\providecommand{\ArXivprefix}{arXiv:}
\providecommand{\URLprefix}{URL: }
\providecommand{\Pubmedprefix}{pmid:}
\providecommand{\doi}[1]{\href{http://dx.doi.org/#1}{\path{#1}}}
\providecommand{\Pubmed}[1]{\href{pmid:#1}{\path{#1}}}
\providecommand{\bibinfo}[2]{#2}
\ifx\xfnm\relax \def\xfnm[#1]{\unskip,\space#1}\fi
%Type = Article
\bibitem[{Arrhenius(1889)}]{Arrhenius1889}
\bibinfo{author}{Arrhenius, S.}, \bibinfo{year}{1889}.
\newblock \bibinfo{title}{Über die reaktionsgeschwindigkeit bei der inversion
  von rohrzucker durch säuren}.
\newblock \bibinfo{journal}{Zeitschrift für Physikalische Chemie}
  \bibinfo{volume}{4}, \bibinfo{pages}{226--248}.
\newblock \URLprefix
  \url{https://www.degruyter.com/view/journals/zpch/4U/1/article-p226.xml}.
%Type = Article
\bibitem[{Carlson(1990)}]{Carlson1990model}
\bibinfo{author}{Carlson, W.D.}, \bibinfo{year}{1990}.
\newblock \bibinfo{title}{Mechanisms and kinetics of apatite fission-track
  annealing}.
\newblock \bibinfo{journal}{American Mineralogist} \bibinfo{volume}{75},
  \bibinfo{pages}{1120--1139}.
\newblock \URLprefix \url{<Go to ISI>://WOS:A1990EL25100017}.
  \bibinfo{note}{times Cited: 116 Carlson, wd Carlson, William/A-5807-2008
  Carlson, William/0000-0002-2954-5886 121}.
%Type = Article
\bibitem[{Carlson et~al.(1999)Carlson, Donelick and Ketcham}]{Carlson1999data}
\bibinfo{author}{Carlson, W.D.}, \bibinfo{author}{Donelick, R.A.},
  \bibinfo{author}{Ketcham, R.A.}, \bibinfo{year}{1999}.
\newblock \bibinfo{title}{Variability of apatite fission-track annealing
  kinetics: I. experimental results}.
\newblock \bibinfo{journal}{American Mineralogist} \bibinfo{volume}{84},
  \bibinfo{pages}{1213--1223}.
\newblock \URLprefix \url{<Go to ISI>://WOS:000082349700001}.
  \bibinfo{note}{times Cited: 346 Carlson, WD Donelick, RA Ketcham, RA Ketcham,
  Richard/B-5431-2011; Carlson, William/A-5807-2008 Ketcham,
  Richard/0000-0002-2748-0409; Carlson, William/0000-0002-2954-5886 370}.
%Type = Article
\bibitem[{Cohen et~al.(2007)Cohen, Cvitas, Fry et~al.}]{cohen2007iupac}
\bibinfo{author}{Cohen, E.}, \bibinfo{author}{Cvitas, T.},
  \bibinfo{author}{Fry, J.}, et~al., \bibinfo{year}{2007}.
\newblock \bibinfo{title}{Iupac quantities, units and symbols in physical
  chemistry}.
\newblock \bibinfo{journal}{IUPAC and RSC Publishing, Cambridge} .
%Type = Article
\bibitem[{C{\'o}rdoba-Torres et~al.(2003)C{\'o}rdoba-Torres, Nogueira and
  Fair{\'e}n}]{cordoba2003fractional}
\bibinfo{author}{C{\'o}rdoba-Torres, P.}, \bibinfo{author}{Nogueira, R.P.},
  \bibinfo{author}{Fair{\'e}n, V.}, \bibinfo{year}{2003}.
\newblock \bibinfo{title}{Fractional reaction order kinetics in electrochemical
  systems involving single-reactant, bimolecular desorption reactions}.
\newblock \bibinfo{journal}{Journal of Electroanalytical Chemistry}
  \bibinfo{volume}{560}, \bibinfo{pages}{25--33}.
%Type = Article
\bibitem[{Crowley et~al.(1991)Crowley, Cameron and
  Schaefer}]{crowley1991Arrhenius}
\bibinfo{author}{Crowley, K.D.}, \bibinfo{author}{Cameron, M.},
  \bibinfo{author}{Schaefer, R.L.}, \bibinfo{year}{1991}.
\newblock \bibinfo{title}{Experimental studies of annealing of etched fission
  tracks in fluorapatite}.
\newblock \bibinfo{journal}{Geochimica Et Cosmochimica Acta}
  \bibinfo{volume}{55}, \bibinfo{pages}{1449--1465}.
\newblock \URLprefix \url{<Go to ISI>://WOS:A1991FM71600019},
  \DOIprefix\doi{10.1016/0016-7037(91)90320-5}. \bibinfo{note}{times Cited: 187
  Crowley, kd cameron, m schaefer, rl 210}.
%Type = Article
\bibitem[{Dodson(1973)}]{dodson1973closure}
\bibinfo{author}{Dodson, M.H.}, \bibinfo{year}{1973}.
\newblock \bibinfo{title}{Closure temperature in cooling geochronological and
  petrological systems}.
\newblock \bibinfo{journal}{Contributions to Mineralogy and Petrology}
  \bibinfo{volume}{40}, \bibinfo{pages}{259--274}.
%Type = Article
\bibitem[{Duddy et~al.(1988)Duddy, Green and Laslett}]{duddy1988thermal}
\bibinfo{author}{Duddy, I.}, \bibinfo{author}{Green, P.},
  \bibinfo{author}{Laslett, G.}, \bibinfo{year}{1988}.
\newblock \bibinfo{title}{Thermal annealing of fission tracks in apatite 3.
  variable temperature behaviour}.
\newblock \bibinfo{journal}{Chemical Geology: Isotope Geoscience section}
  \bibinfo{volume}{73}, \bibinfo{pages}{25--38}.
%Type = Manual
\bibitem[{Elzhov et~al.(2016)Elzhov, Mullen, Spiess and Bolker}]{Elzhov2016}
\bibinfo{author}{Elzhov, T.V.}, \bibinfo{author}{Mullen, K.M.},
  \bibinfo{author}{Spiess, A.N.}, \bibinfo{author}{Bolker, B.},
  \bibinfo{year}{2016}.
\newblock \bibinfo{title}{minpack.lm: R Interface to the Levenberg-Marquardt
  Nonlinear Least-Squares Algorithm Found in MINPACK, Plus Support for Bounds}.
\newblock \URLprefix \url{https://CRAN.R-project.org/package=minpack.lm}.
  \bibinfo{note}{r package version 1.2-1}.
%Type = Article
\bibitem[{Gallagher(2012)}]{gallager2012software}
\bibinfo{author}{Gallagher, K.}, \bibinfo{year}{2012}.
\newblock \bibinfo{title}{Transdimensional inverse thermal history modeling for
  quantitative thermochronology}.
\newblock \bibinfo{journal}{Journal of Geophysical Research-Solid Earth}
  \bibinfo{volume}{117}.
\newblock \URLprefix \url{<Go to ISI>://WOS:000301136400002},
  \DOIprefix\doi{10.1029/2011jb008825}.
%Type = Article
\bibitem[{Goswami et~al.(1984)Goswami, Jha and Lal}]{goswami1984quantitative}
\bibinfo{author}{Goswami, J.}, \bibinfo{author}{Jha, R.}, \bibinfo{author}{Lal,
  D.}, \bibinfo{year}{1984}.
\newblock \bibinfo{title}{Quantitative treatment of annealing of charged
  particle tracks in common minerals}.
\newblock \bibinfo{journal}{Earth and Planetary Science Letters}
  \bibinfo{volume}{71}, \bibinfo{pages}{120--128}.
%Type = Article
\bibitem[{Green et~al.(1988)Green, Duddy and Laslett}]{green1988can}
\bibinfo{author}{Green, P.}, \bibinfo{author}{Duddy, I.},
  \bibinfo{author}{Laslett, G.}, \bibinfo{year}{1988}.
\newblock \bibinfo{title}{Can fission track annealing in apatite be described
  by first-order kinetics?}
\newblock \bibinfo{journal}{Earth and Planetary Science Letters}
  \bibinfo{volume}{87}, \bibinfo{pages}{216--228}.
%Type = Article
\bibitem[{Green et~al.(1986)Green, Duddy, Gleadow, Tingate and
  Laslett}]{Green1986}
\bibinfo{author}{Green, P.F.}, \bibinfo{author}{Duddy, I.R.},
  \bibinfo{author}{Gleadow, A.J.W.}, \bibinfo{author}{Tingate, P.R.},
  \bibinfo{author}{Laslett, G.M.}, \bibinfo{year}{1986}.
\newblock \bibinfo{title}{Thermal annealing of fission tracks in apatite .1. a
  qualitative description}.
\newblock \bibinfo{journal}{Chemical Geology} \bibinfo{volume}{59},
  \bibinfo{pages}{237--253}.
\newblock \URLprefix \url{<Go to ISI>://WOS:A1986F612200002},
  \DOIprefix\doi{10.1016/0009-2541(86)90048-3}.
%Type = Article
\bibitem[{Guedes et~al.(2022)Guedes, Lixandr{\~a}o~Filho and
  Hadler}]{guedes2022generalization}
\bibinfo{author}{Guedes, S.}, \bibinfo{author}{Lixandr{\~a}o~Filho, A.L.},
  \bibinfo{author}{Hadler, J.C.}, \bibinfo{year}{2022}.
\newblock \bibinfo{title}{Generalization of the fission-track arrhenius
  annealing equations}.
\newblock \bibinfo{journal}{Mathematical Geosciences} , \bibinfo{pages}{1--20}.
%Type = Article
\bibitem[{Guedes et~al.(2013)Guedes, Moreira, Devanathan, Weber and
  Hadler}]{guedes2013improved}
\bibinfo{author}{Guedes, S.}, \bibinfo{author}{Moreira, P.A.},
  \bibinfo{author}{Devanathan, R.}, \bibinfo{author}{Weber, W.J.},
  \bibinfo{author}{Hadler, J.C.}, \bibinfo{year}{2013}.
\newblock \bibinfo{title}{Improved zircon fission-track annealing model based
  on reevaluation of annealing data}.
\newblock \bibinfo{journal}{Physics and Chemistry of Minerals}
  \bibinfo{volume}{40}, \bibinfo{pages}{93--106}.
%Type = Article
\bibitem[{Guedes et~al.(2006)Guedes, Oliveira, Moreira, Iunes
  et~al.}]{guedes2006kinetic}
\bibinfo{author}{Guedes, S.}, \bibinfo{author}{Oliveira, K.},
  \bibinfo{author}{Moreira, P.}, \bibinfo{author}{Iunes, P.}, et~al.,
  \bibinfo{year}{2006}.
\newblock \bibinfo{title}{Kinetic model for the annealing of fission tracks in
  minerals and its application to apatite}.
\newblock \bibinfo{journal}{Radiation measurements} \bibinfo{volume}{41},
  \bibinfo{pages}{392--398}.
%Type = Article
\bibitem[{Issler(1996)}]{issler1996TA}
\bibinfo{author}{Issler, D.}, \bibinfo{year}{1996}.
\newblock \bibinfo{title}{Optimizing time step size for apatite fission track
  annealing model}.
\newblock \bibinfo{journal}{Computers and Geosciences} \bibinfo{volume}{22},
  \bibinfo{pages}{835--835}.
%Type = Inbook
\bibitem[{Ketcham(2005)}]{ketcham2005software}
\bibinfo{author}{Ketcham, R.A.}, \bibinfo{year}{2005}.
\newblock \bibinfo{title}{Forward and inverse modeling of low-temperature
  thermochronometry data}. volume~\bibinfo{volume}{58} of
  \textit{\bibinfo{series}{Reviews in Mineralogy and Geochemistry}}.
\newblock pp. \bibinfo{pages}{275--314}.
\newblock \URLprefix \url{<Go to ISI>://WOS:000235184200011},
  \DOIprefix\doi{10.2138/rmg.2005.58.11}.
%Type = Inbook
\bibitem[{Ketcham(2019)}]{Ketcham2019}
\bibinfo{author}{Ketcham, R.A.}, \bibinfo{year}{2019}.
\newblock \bibinfo{title}{Fission-Track Annealing: From Geologic Observations
  to Thermal History Modeling}. \bibinfo{publisher}{Springer International
  Publishing}, \bibinfo{address}{Cham}.
\newblock pp. \bibinfo{pages}{49--75}.
\newblock \DOIprefix\doi{10.1007/978-3-319-89421-8\_3}.
%Type = Article
\bibitem[{Ketcham et~al.(2007)Ketcham, Carter, Donelick, Barbarand and
  Hurford}]{ketcham2007improved}
\bibinfo{author}{Ketcham, R.A.}, \bibinfo{author}{Carter, A.},
  \bibinfo{author}{Donelick, R.A.}, \bibinfo{author}{Barbarand, J.},
  \bibinfo{author}{Hurford, A.J.}, \bibinfo{year}{2007}.
\newblock \bibinfo{title}{Improved modeling of fission-track annealing in
  apatite}.
\newblock \bibinfo{journal}{American Mineralogist} \bibinfo{volume}{92},
  \bibinfo{pages}{799--810}.
%Type = Article
\bibitem[{Ketcham et~al.(1999)Ketcham, Donelick and Carlson}]{Ketcham1999}
\bibinfo{author}{Ketcham, R.A.}, \bibinfo{author}{Donelick, R.A.},
  \bibinfo{author}{Carlson, W.D.}, \bibinfo{year}{1999}.
\newblock \bibinfo{title}{Variability of apatite fission-track annealing
  kinetics: Iii. extrapolation to geological time scales}.
\newblock \bibinfo{journal}{American Mineralogist} \bibinfo{volume}{84},
  \bibinfo{pages}{1235--1255}.
\newblock \URLprefix \url{<Go to ISI>://WOS:000082349700003}.
  \bibinfo{note}{times Cited: 495 Ketcham, RA Donelick, RA Carlson, WD Carlson,
  William/A-5807-2008; Ketcham, Richard/B-5431-2011 Carlson,
  William/0000-0002-2954-5886; Ketcham, Richard/0000-0002-2748-0409 592}.
%Type = Article
\bibitem[{Koga et~al.(1992)Koga, Tanaka and
  {\v{S}}est{\'a}k}]{koga1992fractional}
\bibinfo{author}{Koga, N.}, \bibinfo{author}{Tanaka, H.},
  \bibinfo{author}{{\v{S}}est{\'a}k, J.}, \bibinfo{year}{1992}.
\newblock \bibinfo{title}{On the fractional conversion $\alpha$ in the kinetic
  description of solid-state reactions}.
\newblock \bibinfo{journal}{Journal of thermal analysis} \bibinfo{volume}{38},
  \bibinfo{pages}{2553--2557}.
%Type = Article
\bibitem[{Kooij(1893)}]{Kooij1893}
\bibinfo{author}{Kooij, D.M.}, \bibinfo{year}{1893}.
\newblock \bibinfo{title}{Über die zersetzung des gasförmigen
  phosphorwasserstoffs}.
\newblock \bibinfo{journal}{Zeitschrift für Physikalische Chemie}
  \bibinfo{volume}{12}, \bibinfo{pages}{155 -- 161}.
\newblock \URLprefix
  \url{https://www.degruyter.com/view/journals/zpch/12U/1/article-p155.xml},
  \DOIprefix\doi{https://doi.org/10.1515/zpch-1893-1214}.
%Type = Article
\bibitem[{Laidler(1984)}]{Laidler1984Arrhenius}
\bibinfo{author}{Laidler, K.J.}, \bibinfo{year}{1984}.
\newblock \bibinfo{title}{The development of the arrhenius equation}.
\newblock \bibinfo{journal}{Journal of Chemical Education}
  \bibinfo{volume}{61}, \bibinfo{pages}{494--498}.
\newblock \URLprefix \url{<Go to ISI>://WOS:A1984SX06600005},
  \DOIprefix\doi{10.1021/ed061p494}. \bibinfo{note}{times Cited: 514 Laidler,
  kj 527}.
%Type = Book
\bibitem[{Laidler et~al.(1965)Laidler, Keith et~al.}]{laidler1965chemical}
\bibinfo{author}{Laidler, K.J.}, \bibinfo{author}{Keith, J.}, et~al.,
  \bibinfo{year}{1965}.
\newblock \bibinfo{title}{Chemical kinetics}. volume~\bibinfo{volume}{2}.
\newblock \bibinfo{publisher}{McGraw-Hill New York}.
%Type = Article
\bibitem[{Laslett and Galbraith(1996)}]{LaslettGabraith1996}
\bibinfo{author}{Laslett, G.}, \bibinfo{author}{Galbraith, R.},
  \bibinfo{year}{1996}.
\newblock \bibinfo{title}{Statistical modelling of thermal annealing of fission
  tracks in apatite}.
\newblock \bibinfo{journal}{Geochimica et Cosmochimica Acta}
  \bibinfo{volume}{60}, \bibinfo{pages}{5117 -- 5131}.
\newblock \URLprefix
  \url{http://www.sciencedirect.com/science/article/pii/S0016703796003079},
  \DOIprefix\doi{https://doi.org/10.1016/S0016-7037(96)00307-9}.
%Type = Article
\bibitem[{Laslett et~al.(1987)Laslett, Green, Duddy and
  Gleadow}]{laslett1987thermal}
\bibinfo{author}{Laslett, G.}, \bibinfo{author}{Green, P.F.},
  \bibinfo{author}{Duddy, I.}, \bibinfo{author}{Gleadow, A.},
  \bibinfo{year}{1987}.
\newblock \bibinfo{title}{Thermal annealing of fission tracks in apatite 2. a
  quantitative analysis}.
\newblock \bibinfo{journal}{Chemical Geology: Isotope Geoscience Section}
  \bibinfo{volume}{65}, \bibinfo{pages}{1--13}.
%Type = Article
\bibitem[{Rana et~al.(2021)Rana, Lixandrao and Guedes}]{Rana2021}
\bibinfo{author}{Rana, M.A.}, \bibinfo{author}{Lixandrao, A.L.},
  \bibinfo{author}{Guedes, S.}, \bibinfo{year}{2021}.
\newblock \bibinfo{title}{A new phenomenological model for annealing of fission
  tracks in apatite: laboratory data fitting and geological benchmarking}.
\newblock \bibinfo{journal}{Physics and Chemistry of Minerals}
  \bibinfo{volume}{48}.
\newblock \URLprefix \url{<Go to ISI>://WOS:000641849100002},
  \DOIprefix\doi{10.1007/s00269-021-01143-9}.
%Type = Article
\bibitem[{Rufino and Guedes(2022)}]{rufino2022arrhenius}
\bibinfo{author}{Rufino, M.}, \bibinfo{author}{Guedes, S.},
  \bibinfo{year}{2022}.
\newblock \bibinfo{title}{Arrhenius activation energy and transitivity in
  fission-track annealing equations}.
\newblock \bibinfo{journal}{Chemical Geology} , \bibinfo{pages}{120779}.
%Type = Article
\bibitem[{Tamer and Ketcham(2020)}]{tamer2020low}
\bibinfo{author}{Tamer, M.}, \bibinfo{author}{Ketcham, R.},
  \bibinfo{year}{2020}.
\newblock \bibinfo{title}{Is low-temperature fission-track annealing in apatite
  a thermally controlled process?}
\newblock \bibinfo{journal}{Geochemistry, Geophysics, Geosystems}
  \bibinfo{volume}{21}, \bibinfo{pages}{e2019GC008877}.
%Type = Book
\bibitem[{Vyazovkin(2015)}]{Vyazovkin2015Book}
\bibinfo{author}{Vyazovkin, S.}, \bibinfo{year}{2015}.
\newblock \bibinfo{title}{Isoconversional Kinetics of Thermally Stimulated
  Processes}.
\newblock \bibinfo{edition}{1} ed., \bibinfo{publisher}{Springer, Cham}.
\newblock \DOIprefix\doi{https://doi.org/10.1007/978-3-319-14175-6}.
%Type = Article
\bibitem[{Vyazovkin(2016)}]{vyazovkin2016time}
\bibinfo{author}{Vyazovkin, S.}, \bibinfo{year}{2016}.
\newblock \bibinfo{title}{A time to search: finding the meaning of variable
  activation energy}.
\newblock \bibinfo{journal}{Physical Chemistry Chemical Physics}
  \bibinfo{volume}{18}, \bibinfo{pages}{18643--18656}.
%Type = Article
\bibitem[{Wauschkuhn et~al.(2015)Wauschkuhn, Jonckheere and
  Ratschbacher}]{Wauschkuhn2015KTB}
\bibinfo{author}{Wauschkuhn, B.}, \bibinfo{author}{Jonckheere, R.},
  \bibinfo{author}{Ratschbacher, L.}, \bibinfo{year}{2015}.
\newblock \bibinfo{title}{The ktb apatite fission-track profiles: Building on a
  firm foundation?}
\newblock \bibinfo{journal}{Geochimica Et Cosmochimica Acta}
  \bibinfo{volume}{167}, \bibinfo{pages}{27--62}.
\newblock \URLprefix \url{<Go to ISI>://WOS:000361007300003},
  \DOIprefix\doi{10.1016/j.gca.2015.06.015}. \bibinfo{note}{times Cited: 11
  Wauschkuhn, B. Jonckheere, R. Ratschbacher, L. Ratschbacher,
  Lothar/0000-0001-9960-2084; Wauschkuhn, Bastian/0000-0002-4684-5178 11
  1872-9533}.
%Type = Article
\bibitem[{Wolfram-Research-Inc.(2021)}]{Mathematica}
\bibinfo{author}{Wolfram-Research-Inc.}, \bibinfo{year}{2021}.
\newblock \bibinfo{title}{Mathematica, {V}ersion 12.3}.
\newblock \bibinfo{journal}{Champaign, IL} \URLprefix
  \url{https://www.wolfram.com/mathematica}.

\end{thebibliography}

\end{document}